\documentclass[prd,twocolumn,nofootinbib,superscriptaddress]{revtex4-2}

\usepackage{comment} 

\usepackage{graphicx}
\usepackage{amsmath,bm,amssymb,amsfonts,dsfont}
\usepackage[usenames,dvipsnames]{xcolor}
\usepackage[normalem]{ulem}
\usepackage{url}
\usepackage{array}
\usepackage{booktabs}
\usepackage{multirow}
\usepackage{float}
\usepackage[colorlinks  = true,
            linkcolor   = NavyBlue,
            urlcolor    = NavyBlue,
            citecolor   = NavyBlue,
            anchorcolor = NavyBlue]{hyperref}
\usepackage{cprotect}

\usepackage{mathtools,slashed}

\usepackage{appendix}

\usepackage{bbold}

\usepackage[compat=1.1.0]{tikz-feynman} % TikZ-Feynman (TikZ by Till Tantau)
\tikzfeynmanset{every vertex/.style={/tikz/font=\small}}

\begin{document}

\title{Two Puzzles, One Solution: Neutrino Mass and Secluded Dark Matter}

\author{Mattia Di Mauro}\email{dimauro.mattia@gmail.com}
\affiliation{Istituto Nazionale di Fisica Nucleare, Sezione di Torino, Via P. Giuria 1, 10125 Torino, Italy}

\date{\today}

\begin{abstract}
We present a minimal secluded dark-matter (DM) framework based on an extra $U(1)_X$ gauge symmetry. The model contains a Dirac DM particle $\chi$, three heavy neutrinos $N_I$ with masses $M_{N,I}$, and a singlet scalar $R$ that mixes with the Standard Model Higgs doublet $\Phi$ by an angle $\alpha$. A symmetry forbids the $\Phi$-$R$ portal at tree level; the leading portal then arises at one loop from the same Yukawa structures that generate active neutrino masses $m_{\nu,I}$, implying $\tan(2\alpha) \propto \sum_I m_{\nu,I} M^2_{N,I}/(v_h m_H^2)$, where $v_h$ and $m_H$ are the SM Higgs VEV and mass. For heavy-neutrino masses in the multi-TeV range, this yields a naturally tiny mixing, $\tan(2\alpha)\sim 5\times 10^{-11}\left(M_N/10~\mathrm{TeV}\right)^2$, which strongly suppresses DM signals in direct, indirect, and collider searches. For PeV-scale heavy neutrinos the loop-induced portal is enhanced and the DM--nucleon cross section can instead enter the reach of direct-detection experiments. The visible and dark sectors thermalize at temperatures of order a few times the mass of the lightest heavy neutrino, then subsequently decouple, and typically evolve with a slightly hotter dark bath. In the secluded regime, with $\tan(2\alpha)\ll 1$ and $m_\chi>m_{H_p}$, the relic density is set by $p$-wave annihilation $\chi\bar\chi \to H_p H_p$ (with $H_p$ the Higgs-like particle of the dark sector), and the dark-sector Yukawa couplings required to reproduce the observed abundance are $\mathcal{O}(0.1\text{--}1)$, as in the standard WIMP case. For heavy-neutrino masses $\gtrsim 10~\mathrm{TeV}$, the mediator decays before nucleosynthesis without spoiling BBN observables, while the tiny portal suppresses present-day signals below current and near-future sensitivities. This links two long-standing puzzles -- the absence of DM signals and the smallness of neutrino masses -- within a predictive thermal framework.
\end{abstract}

\maketitle

{\it Introduction.}
The existence of dark matter (DM) is firmly established through its gravitational effects, yet no laboratory experiment has detected any signal of DM–Standard Model (SM) interactions to date~\cite{Bertone:2010zza,Bertone:2016nfn,Cirelli:2024ssz}. A viable DM candidate must be stable, neutral, non-relativistic at matter–radiation equality, and weakly interacting~\cite{Bertone:2004pz,Cirelli:2024ssz}. WIMPs constitute a well-motivated class of DM candidates whose thermal freeze-out at electroweak scales naturally yields \(\Omega_{\rm DM} h^2 \simeq 0.12\)~\cite{Lee:1977ua,1978ApJ...223.1015G,WESS197439,1983PhRvL..50.1419G,Ellis:1983ew,Steigman:1984ac}, motivating extensive direct, collider, and indirect searches~\cite{Schumann:2019eaa,Boveia:2018yeb,Gaskins:2016cha} alongside cosmological observations~\cite{Aghanim:2018eyx}.

Current limits, e.g., LZ and XENONnT, probe spin–independent cross sections down to \(\mathcal{O}(10^{-47}\text{--}10^{-48})\,\mathrm{cm}^2\) for weak-scale masses~\cite{LZ:2023,Aprile:2023XENONnT,LZ:2024zvo}, excluding broad regions of parameter space for models where the same portal controls both freeze-out and scattering~\cite{Arcadi:2017kky,Arcadi:2019lka,DiMauro:2023tho,Arcadi:2024ukq,DiMauro:2025jia}. Resonant annihilation can evade these bounds, though typically at the price of mass tuning~\cite{DiMauro:2025jia,DiMauro:2023tho}. A more generic alternative is \emph{secluded} DM~\cite{Pospelov:2007mp,Pospelov:2008jd,DiMauro:2025jsb}, in which the relic density is set by DM annihilating into mediator pairs within a dark sector while the SM portal (arising from Higgs mixing \(\sin\alpha\) or gauge kinetic mixing) is parameterized by a coupling \(\epsilon\), which typically must satisfy \(\epsilon \ll 10^{-3}\). Such a tiny portal might at first seem fine-tuned and theoretically unmotivated.

Interestingly, the SM already contains another unusually small mass scale: the active neutrino mass \(m_\nu\), with possible values in the tens-of-meV range~\cite{Planck2018,DESI:2024VI,NuFIT:6.0,KATRIN:2024cdt}, i.e., \(\sim 10^{-7}\) of the electron mass.

We propose that the required smallness of the SM–dark portal is \emph{not} fine-tuned but can be naturally connected to the small neutrino masses. In a minimal setup where a symmetry forbids the tree-level portal, the leading SM–dark coupling is generated radiatively by the same Yukawa structures that yield neutrino masses via a type-I seesaw~\cite{Minkowski:1977sc,Yanagida:1979as,GellMann:1980vs,Mohapatra:1979ia,Wyler:1982dd,Mohapatra:1986aw}. The induced portal—either a Higgs mixing angle \(\sin\alpha\) or a gauge kinetic mixing \(\epsilon\)—then scales with \(m_\nu\) up to loop factors and mass ratios. Consequently, direct, indirect, and collider signals are naturally suppressed—because \(m_\nu\) is tiny—while the relic abundance remains set by DM annihilations into dark mediator pairs and is largely decoupled from the portal.
%Recent data placing neutrino masses at the tens-of-meV scale~\cite{Planck2018,DESI:2024VI,NuFIT:6.0,KATRIN:2024cdt} make this connection quantitatively compelling.

{\it Model Lagrangian and particle content.}
We consider a minimal BSM dark-sector setup that is UV-complete and can realize the secluded DM mechanism. We introduce an extra abelian gauge group
$U(1)_X$ containing a Dirac fermion $\chi$ with charge $q_X^\chi$, singlet under
the SM gauge groups, a massive gauge boson $Z'$ with coupling $g_X$, three
right-handed neutrinos $N_i$ ($i=1,2,3$), and a complex scalar singlet $R$
that mixes with the SM Higgs doublet $\Phi$ (see
e.g.~\cite{Baek:2011aa,Baek:2013qwa}). We report all the details in App.~\ref{sec:model} and summarize here its main features. The relevant
Lagrangian terms are
\begin{widetext}
\begin{eqnarray}
\mathcal{L} \!\supset \!
- \frac{1}{4} F'_{\mu\nu}F'^{\mu\nu}\!\!\!
-\!\frac{\epsilon}{2}\, F'_{\mu\nu} B^{\mu\nu}\!
+\! \partial_\mu R^\dagger \partial^\mu R \!
-\! V(\Phi,R)
+ \bar\chi\,(i\!\not\!\! D - m_\chi)\chi
\!-\!\Big( y_p \bar\chi\chi R +
Y_\nu^{\alpha i}\,\bar L_\alpha\tilde{\Phi} N_i
\!+\! \tfrac12 Y_N^{ij}\,R\,\overline{N_i^{\;c}}N_j + \text{h.c.}\!\Big),
\label{eq:Ltot}
\end{eqnarray}
\end{widetext}
where $L_\alpha=(\nu_{\alpha L},\ell_{\alpha L})^T$ are the SM lepton doublets, $\Phi$ is the SM Higgs, $B_{\mu\nu}$ is the hypercharge field-strength tensor,
$Y_\nu$ and $Y_N=Y_N^T$ are the generic complex $3\times3$ Yukawa matrices coupling $\Phi$ and $R$ to the light and heavy neutrinos.
We take $\chi$ to be vector-like under $U(1)_X$, i.e., $q^{\chi}_{L}=q^{\chi}_{R}=q^\chi_X$.
Then a bare Dirac mass term $m_0 \bar\chi\chi$ is gauge invariant if $R$ is
neutral ($q_X^R=0$), and we can add a renormalizable Yukawa portal
$-y_p \bar\chi\chi R$. In this case $U(1)_X$ is not broken by $R$ acquiring a vacuum expectation value (VEV) $v_r=\langle R \rangle$, and the $Z'$ mass can be generated
via a Stückelberg mechanism. Since $\chi$ is vector-like, $y_p$ is not fixed by $m_\chi$
($m_\chi = m_0 + y_p v_r/\sqrt{2}$).
Here $F'_{\mu\nu}$, $g_X$, and $m_{Z'}$ denote the field strength, coupling, and mass
of the new gauge boson $Z'_\mu$. The term
$-\frac{\epsilon}{2}\, F'_{\mu\nu} B^{\mu\nu}$ represents kinetic mixing between
$U(1)_X$ and hypercharge, controlled by $\epsilon$~\cite{Holdom:1985ag,Essig:2013lka}.
In the following we neglect this mixing (or assume it is sufficiently small). SM fermions are neutral under $U(1)_X$, so they do not couple directly to $Z'$ at tree level.

The $(R,\Phi)$ scalar potential is
\begin{eqnarray}
V(\Phi,R)
&=& \mu_H^2\,\Phi^\dagger\Phi + \mu_R^2\, R^\dagger R
+\lambda_H (\Phi^\dagger\Phi)^2 +  \nonumber \\
&+&\lambda_R (R^\dagger R)^2
+\kappa (\Phi^\dagger\Phi)(R^\dagger R)\,.
\end{eqnarray}
After electroweak symmetry breaking (EWSB) $\Phi \to (v_h+h)/\sqrt{2}$ and $R \to (v_r+\rho)/\sqrt{2}$,
the CP-even mass matrix in the $(h,\rho)$ basis is
\begin{equation}
\mathcal{M}^2=
\begin{pmatrix}
2\lambda_H v_h^2 & \kappa v_h v_r \\
\kappa v_h v_r & 2\lambda_R v_r^2
\end{pmatrix},
\,\,\,
\tan 2\alpha = \frac{\kappa \, v_h v_r}{\lambda_R v_r^2 - \lambda_H v_h^2}\,,
\label{eq:mixingp}
\end{equation}
which is diagonalized by a rotation of angle $\alpha$, yielding the SM-like state $H$
and the singlet-like state $H_p$ with masses $m_H$ and $m_{H_p}$.

%The couplings of $H,H_p$ to DM and SM fermions follow from this rotation:
%\begin{align}
%g_{\chi\chi H}      &= \frac{y_p}{\sqrt{2}}\,\sin\alpha\,, &
%g_{\chi\chi H_p}    &= \frac{y_p}{\sqrt{2}}\,\cos\alpha\,, \\
%g_{H f\bar f}       &= \frac{m_f}{v_h}\,\cos\alpha\,, &
%g_{H_p f\bar f}     &= -\,\frac{m_f}{v_h}\,\sin\alpha\,,
%\end{align}
%and $Z'$ couples to $\chi$ with strength $g_\chi = g_X q_\chi$. 

\medskip

{\it Type-I seesaw.}
We summarize the main aspects of light and heavy neutrino masses, couplings, and the seesaw mechanism; a more detailed discussion is given in App.~\ref{sec:neutrino}.
Taking the heavy Majorana mass matrix to be
$M_N = \tfrac{v_r}{\sqrt{2}}\,Y_N$ and the Dirac mass matrix
$m_D = \tfrac{v_h}{\sqrt{2}}\,Y_\nu$, the Majorana-like neutrino mass matrix in the
$(\nu_L,\,N^{\,c})$ basis is
\begin{equation}
\mathcal{M}_\nu=
\begin{pmatrix}
0 & m_D\\
m_D^T & M_N
\end{pmatrix}.
\end{equation}
For $\|m_D\|\ll \|M_N\|$ one obtains the standard type-I seesaw relation
\begin{equation}
M_\nu \;\simeq\; -\,m_D\,M_N^{-1}\,m_D^T.
\end{equation}
Since $Y_N$ is symmetric, $M_N$ is diagonalized by a Takagi factorization, defining the heavy–Majorana–neutrino mass eigenstates $N_I$ with masses $M_{N,I}>0$ ($I=1,2,3$) and diagonal Yukawa entries $y_{N,I}= \sqrt{2}\,M_{N,I}/v_r$. The light–neutrino mass matrix $M_\nu$ is diagonalized by the PMNS matrix $U_\nu$, yielding the physical light eigenstates $\nu_i$ with masses $m_i$. The Majorana mass term breaks the global lepton number symmetry $U(1)_L$ by two units ($\Delta L=2$), so the light neutrinos are Majorana.

In this letter we work in the basis where the heavy–neutrino Majorana mass matrix is diagonal, and we parameterize the Dirac mass matrix via the Casas–Ibarra form~\cite{Casas:2001sr}:
\begin{equation}
m_D = i\,U_\nu\,\sqrt{\widehat m_\nu}\,O\,\sqrt{\widehat M_N},
\end{equation}
where $\widehat m_\nu = \mathrm{diag}(m_{\nu,1},m_{\nu,2},m_{\nu,3})$ and $O$ is a complex orthogonal matrix, $O^T O = \mathbb{1}$. In the \emph{aligned limit} we take $O=\mathbb{1}$ and use the freedom to label the heavy eigenstates so that each heavy mass $M_{N,I}$ is paired with the corresponding light eigenvalue $m_{\nu,I}$. In this case,
\begin{equation}
(Y_\nu^\dagger Y_\nu)_{II}
= \sum_{\alpha=e,\mu,\tau} |y_{\alpha I}|^2
= \frac{2\,m_{\nu,I}\,M_{N,I}}{v_h^2}.
\label{eq:YdagY-aligned}
\end{equation}
The usual effective seesaw mass parameter for $N_I$ becomes $m_{\nu,I}$,
\begin{equation}
\widetilde m_I
\equiv \frac{(m_D^\dagger m_D)_{II}}{M_{N,I}} = m_{\nu,I}
= \frac{v_h^2}{2}\,\frac{|y_{\nu,I}|^2}{M_{N,I}}.
\label{eq:mnuI}
\end{equation}
where $y_{\nu,I}$ are the Yukawa couplings associated to $m_{\nu,I}$ as defined in Eq.~\ref{eq:YdagY-aligned}.

%The couplings present in the vertex $n_i N_I H$ ($n_i N_I H_p$) is proportional to $\cos{\alpha}$ ($\sin{\alpha}$). Instead, the vertex $N_I N_I H$ ($N_I N_I H_p$) is proportional to $\cos{\alpha}$ ($\sin{\alpha}$).

%The interaction between the massive neutrinos and the dark scalar $R$ is present in the lagrangian term $\tfrac12 y_N^{ij}\,R\,\overline{N_i^{\;c}}N_j$.
%The Yukawa couplings of the neutrinos to the SM and dark Higgs are parametrized throght the parameters: $y_\nu^{\alpha i}$ that couples the Higgs doublet to left- and right-handed neutrinos and $y_N^{ij}$ (symmetric in \(i,j\)) that couples the singlet $R$ to
%the Majorana pair \(N_i N_j\).
%After EWSB and dark breaking, the type-I seesaw mechanism provides the masses to the neutrinos,
%\begin{equation}
% m_D = y_\nu\frac{v}{\sqrt2},\, M_N = y_N\frac{v_\Phi}{\sqrt2},\, m_\nu \simeq - m_D^T M_N^{-1} m_D,
% \label{eq:seesaw}
%\end{equation}
%where $m_D$ is the Dirac mass matrix, $M_N$ the Majorana mass matrix and $m_\nu$ is the mass of the light neutrino.

{\it WIMP and secluded regimes.}
The WIMP regime applies for $m_\chi < m_{H_p}$ with $y_p\sim \mathcal{O}(1)$ and $\sin{\alpha} \sim \mathcal{O}(0.1)$. In this case, the dominant annihilation channels are $\chi\bar\chi \to f\bar f$ through $s$–channel exchange of $H$ or $H_p$, with
\begin{equation}
    \langle \sigma v \rangle_{f\bar f} \propto y_p^2 \big(\sin\alpha \cos\alpha\big)^2.
\label{eq:annff}
\end{equation}
In this regime DM can reach the correct relic density for $y_p\sim 0.1$–$1$, but these couplings are almost entirely ruled out by direct–detection limits (except near the resonance $m_\chi \simeq m_{H_p}/2$), see Ref.~\cite{DiMauro:2025jsb}.

In the secluded case, valid for $m_\chi > m_{H_p}$ with $y_p \sim \mathcal{O}(0.1)$ and $\sin{\alpha}\ll 1$, annihilations into dark–sector states dominate, e.g.\ $\chi\bar\chi \to H_p H_p$, with
\begin{equation}
    \langle \sigma v \rangle_{H_p H_p} \propto \frac{v_{\rm rel}^2\,y_p^4}{m_\chi^2},
\end{equation}
where $v_{\rm{rel}}$ is the DM relative velocity.
Spin–independent scattering on nucleons proceeds via $t$–channel $H$ or $H_p$ exchange and scales as
\begin{equation}
    \sigma^{\rm SI}_{\chi N} \propto y_p^2 \cos^{2}\alpha \,\sin^{2}\alpha \;\propto\; y_p^2 \big(\tan{2\alpha}\big)^2,
\end{equation}
so in the secluded regime the nuclear cross section and indirect detection (see Eq.~\ref{eq:annff}) are highly suppressed ($\tan{2\alpha}\ll1$), while for $y_p\sim \mathcal{O}(0.1$–$1)$ the process $\chi\bar\chi\to H_p H_p$ yields $\Omega_{\rm DM} h^2 \simeq 0.12$. Throughout, since we work in the secluded case, we use $\tan{2\alpha}$ to characterize the mixing strength between $\Phi$ and $R$.

{\it Tree-level portals set to zero.}
The scalar portal $|R|^2|\Phi|^2$ (a similar reasoning can be applied to the abelian kinetic mixing $\epsilon\,F^{\mu\nu}_Y F^D_{\mu\nu}$) is gauge–invariant and thus generic in 4D renormalizable EFTs \cite{Holdom:1985ag,Dienes:1996zr}. We therefore \emph{impose} $\kappa(\Lambda)=0$ at the UV scale ($\Lambda$), and arrange that the only fields communicating between the SM and the dark sector are the neutrino-sector spurions (the mass matrix $M_N$) that also generate the light neutrino masses $m_{\nu,I}$. This makes any nonzero portal \emph{radiative} and aligned with neutrino-mass breaking.
If $\kappa(\Lambda)=0$ at tree level, the running of the portal at lower energies also remains zero. This is due to the fact that the function $\beta_{\kappa}=d\kappa/d\log{\mu}$ does not have any additive terms (see App.~\ref{app:kappa-running}).

Two standard UV mechanisms that ensure a vanishing tree-level portal are technically natural:
\textbf{(i) Sequestering of sectors.} In extra-dimensional or sequestered SUSY setups the renormalizable Lagrangian factorizes, so $|H|^2|\Phi|^2$ is absent at tree level and is induced only by messengers in the loops (here, the $N_i$) \cite{ArkaniHamed:1998rs,Randall:1999ee,Luty:2001zv,Randall:1998uk}.
\textbf{(ii) Gauge-portal protection.} For kinetic mixing, either a non-abelian UV origin $G_D\!\supset\!U(1)_D$ forbids $F_YF_D$ until symmetry breaking \cite{Dienes:1996zr,Holdom:1985ag}, or a dark charge-conjugation $C_D$ makes $F_YF_D$ odd, so $\epsilon$ first appears with the same spurions that generate $m_\nu$ \cite{Feldman:2007wj,Pospelov:2008jd}.

%In all such cases, $\kappa(\Lambda)=0,\, \epsilon(\Lambda)=0$, and the leading contributions are loop-induced as we will show later in the letter. In the ’t~Hooft sense \cite{tHooft:1979rat} this is natural: as the neutrino spurions $M_N$ are switched off, lepton number is restored and both $m_\nu$ and the portals vanish.

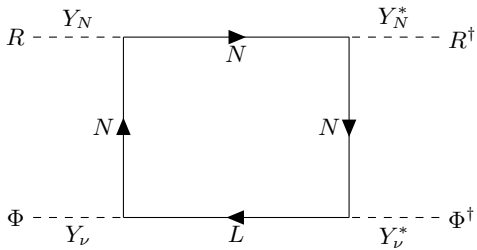
\begin{figure}
\centering

% Full one-loop box
\begin{tikzpicture}[baseline={(current bounding box.center)}]
\begin{feynman}
  % corners of the loop (clockwise)
  \vertex (A) at (0,  1.2);
  \vertex (B) at (3.0, 1.2);
  \vertex (C) at (3.0,-1.2);
  \vertex (D) at (0, -1.2);

  % external scalars
  \vertex[left=1.2cm of A] (PhiL) {$R$};
  \vertex[right=1.2cm of B] (PhiR) {$R^\dagger$};
  \vertex[right=1.2cm of C] (HbotR) {$\Phi^\dagger$};
  \vertex[left=1.2cm of D] (HbotL) {$\Phi$};

  \diagram*{
    (A) -- [fermion, edge label'=$N$] (B)
        -- [fermion, edge label'=$N$] (C)
        -- [fermion, edge label=$L$]   (D)
        -- [fermion, edge label=$N$]   (A),

    (PhiL)  -- [scalar, edge label=$Y_N$] (A),
    (PhiR)  -- [scalar, edge label'=$Y_N^\ast$] (B),
    (HbotR) -- [scalar, edge label=$Y_\nu^\ast$] (C),
    (HbotL) -- [scalar, edge label'=$Y_\nu$] (D),
  };
\end{feynman}
\end{tikzpicture}
\hspace{2.8cm}
% EFT contact operator
\caption{Box Feynman diagram describing the one-loop generation of the mixed quartic $(R^\dagger R)(\Phi^\dagger \Phi)$.}
\label{fig:correct-box-paper}
\end{figure}

{\it Loop–induced portal from the neutrino sector.}
%In our setup the tree–level operator \(\kappa_0\,(\Phi^\dagger \Phi)(R^\dagger R)\) is assumed absent, so the leading interaction between the SM Higgs doublet \(\Phi\) and the dark scalar \(R\) is generated radiatively by the fields that connect the two sectors, namely the lepton doublets \(L_\alpha\) and the heavy singlet neutrinos \(N_I\).
At one loop an effective vertex \(\kappa_{\rm{loop}}(\Phi^\dagger \Phi)(R^\dagger R)\) arises from a box diagram with two \(\Phi L N\) and two \(R N N\) insertions
(see Fig.~\ref{fig:correct-box-paper}).
Considering three heavy singlets \(N_I\) and working in the aligned
Casas–Ibarra limit for the Dirac mass matrix, in dimensional regularization
($\overline{\text{MS}}$) and matching at the renormalization scale $\mu = M_{N}$, the loop–induced coupling is
\begin{equation}
\kappa_{\rm loop}
=
-\sum_{I=1}^3
\frac{y_{N,I}^2\,M_{N,I}}{8\pi^2 v_h^2}\;m_{\nu,I}\,,
\label{eq:kloop}
\end{equation}
which makes explicit the parametric correlation with the light neutrino masses \(\kappa_{\rm loop}\propto \sum_Im_{\nu,I}\).
%For a general complex–orthogonal matrix \(O\) in the Casas–Ibarra parameterization the numerical coefficients change by \(\mathcal{O}(1)\) factors, but the scaling with neutrino masses and Yukawas remains.
For momenta \(|p^2|\!\ll\! M_{N,I}^2\), the one–loop amplitude matches onto the local operator \((\Phi^\dagger\Phi)(R^\dagger R)\), so we use \(\kappa_{\rm loop}\) as the effective \(\Phi\Phi RR\) contact coupling that is valid for freeze–out, direct detection, and the collider observables.
%The running of $\kappa_{\rm loop}$ from the scale $\Lambda$ to the lower energies can change at most of a factor of a few. Therefore, the main conclusions of the letter are not affected by the evolution of $\kappa_{\rm loop}(\mu)$.

{\it Relation between the mixing angle and $m_\nu$.}
At tree level the scalar mixing angle $\alpha$ between the SM-like Higgs $H$ and the singlet-like state $H_p$ is controlled by the portal coupling $\kappa$ via Eq.~\eqref{eq:mixingp}. 
%In terms of physical masses,
%\begin{equation}
%\tan 2\alpha
%=
%\frac{\kappa\,v_h v_r}{\lambda_R v_r^2 - \lambda_H v_h^2}
%=
%\frac{2\,\kappa\,v_h v_r}{m_{H_p}^2 - m_H^2}\,.
%\label{eq:mixing-app}
%\end{equation}
Using Eq.~\eqref{eq:mixingp} with $\kappa=\kappa_{\rm loop}$ from Eq.~\eqref{eq:kloop} and $M_N=y_N v_r/\sqrt{2}$, one finds for a single generation
\begin{eqnarray}
&\tan 2\alpha
=
-\,\frac{|m_\nu|\,y_N\,M_N^2}{2\sqrt{2}\,\pi^2\,v_h\,(m_{H_p}^2 - m_H^2)} \\
&\simeq \!
-\frac{4.66\times 10^{-11}}{1 - (m_{H_p}/125~\mathrm{GeV})^2}\;
y_N \!
\left(\frac{|m_\nu|}{0.05~\mathrm{eV}}\right)\!
\left(\frac{M_N}{10~\mathrm{TeV}}\right)^{\!2}.\nonumber
\label{eq:tan2a-1gen}
\end{eqnarray}
For three generations (e.g.~in the Casas--Ibarra aligned limit) the result generalizes by
$y_N M_N^2 |m_\nu| \to \sum_I y_{N,I}\,M_{N,I}^2\,|m_{\nu,I}|$.

Therefore, the $H$–$H_p$ mixing that controls the SM–dark sector portal is directly proportional to the light neutrino masses. The smallness of $m_\nu$ (from the seesaw mechanism) explains the small portal. In particular, values $\tan 2\alpha \sim 10^{-11}$ render direct, indirect, and collider signals essentially undetectable with current and next-generation experiments (see \cite{DiMauro:2025jsb}).
%e.g.\ the spin–independent rate is suppressed as $\sigma_{\rm SI}\propto y_p^2\sin^2\alpha$, placing it roughly $5$–$6$ orders of magnitude below the neutrino floor.
%working in the heavy-neutrino mass basis with $M_N = \mathrm{diag}(M_1,M_2,M_3)$ and using the Casas--Ibarra parametrization $m_D = i\,U_\nu\,\sqrt{\widehat m_\nu}\,O\,\sqrt{\widehat M_N}$ with $O = \mathbb{1}$, the loop-induced portal is governed by $(Y_\nu^\dagger Y_\nu)_{II}$ and $y_{N,I}$, and the mixing angle generalises to
%\begin{widetext}
%\begin{equation}
%\tan 2\alpha
%\simeq
%-\frac{2.33\times10^{-11}}{1 - (m_{H_p}/125~\mathrm{GeV})^2}
%\sum_{I=1}^3
%\left[
%y_{N,I}\;
%\left(\frac{ m_{\nu,I}}{0.05~\mathrm{eV}}\right)
%\left(\frac{M_I}{10~\mathrm{TeV}}\right)^{\!2}
%\right],
%\label{eq:tan2a-3gen}
%\end{equation}
%\end{widetext}
%where $\widehat m_\nu = \mathrm{diag}(m_{1,\nu},m_{2,\nu},m_{3,\nu})$ and $O$
%is a (complex) orthogonal matrix. In words: once the tree-level
%$\Phi$--$R$ portal is forbidden, the scalar mixing angle is generated
%radiatively and becomes parametrically proportional to the light-neutrino
%masses (up to loop factors and ratios of heavy scales).

{\it Dark matter relic density.}
Here we detail how DM attains the observed relic density, following the evolution of the SM and hidden sectors from very high temperatures, when the heavy neutrinos are ultra-relativistic ($T\gg M_N$), down to the BBN epoch at $t_{\rm BBN}\sim 1~\text{s}$.

We start from the end, i.e.~the BBN epoch. We focus on the secluded regime with $m_\chi>m_{H_p}$ and $\tan(2\alpha)\ll 1$, so that $\chi\chi\to H_p H_p$ controls the relic abundance. After DM freezes out in the hidden sector, the produced $H_p$ remains in the dark bath and subsequently decays into SM fermions; if these decays occur after BBN, they can spoil light-element abundances.

Neglecting threshold effects and summing over open fermionic channels for $m_{H_p}\in[20,100]~\text{GeV}$, a convenient parametrization of the $H_p$ lifetime is
\begin{equation}
\tau_{\rm ferm}(H_p)
=
\frac{1}{\Gamma_{\rm ferm}}
\;\approx\;
4.3\times10^{-21}\,{\rm s}\,
\left(\frac{20~{\rm GeV}}{m_{H_p}}\right)
\frac{1}{\big(\tan 2\alpha\big)^2},
\end{equation}
up to $\mathcal{O}(1)$ factors from channel thresholds. Using the loop-induced mixing and $\tan{2\alpha}$ relation of Eq.~\eqref{eq:tan2a-1gen} one obtains for one generation
\begin{widetext}
\begin{equation}
\tau_{\rm ferm}(H_p)
\simeq
\left(\frac{1.20~\mathrm{s}}{y_N^2}\right)
\left(\frac{20~\mathrm{GeV}}{m_{H_p}}\right)
\left(\frac{16~\mathrm{TeV}}{M_N}\right)^{4}
\left(\frac{0.05~\mathrm{eV}}{m_\nu}\right)^{2}
\left[1-\left(\frac{m_{H_p}}{125~\mathrm{GeV}}\right)^2\right]^2.
\label{eq:GammaF_simple}
\end{equation}
\end{widetext}

\begin{figure}
\includegraphics[width=0.99\linewidth]{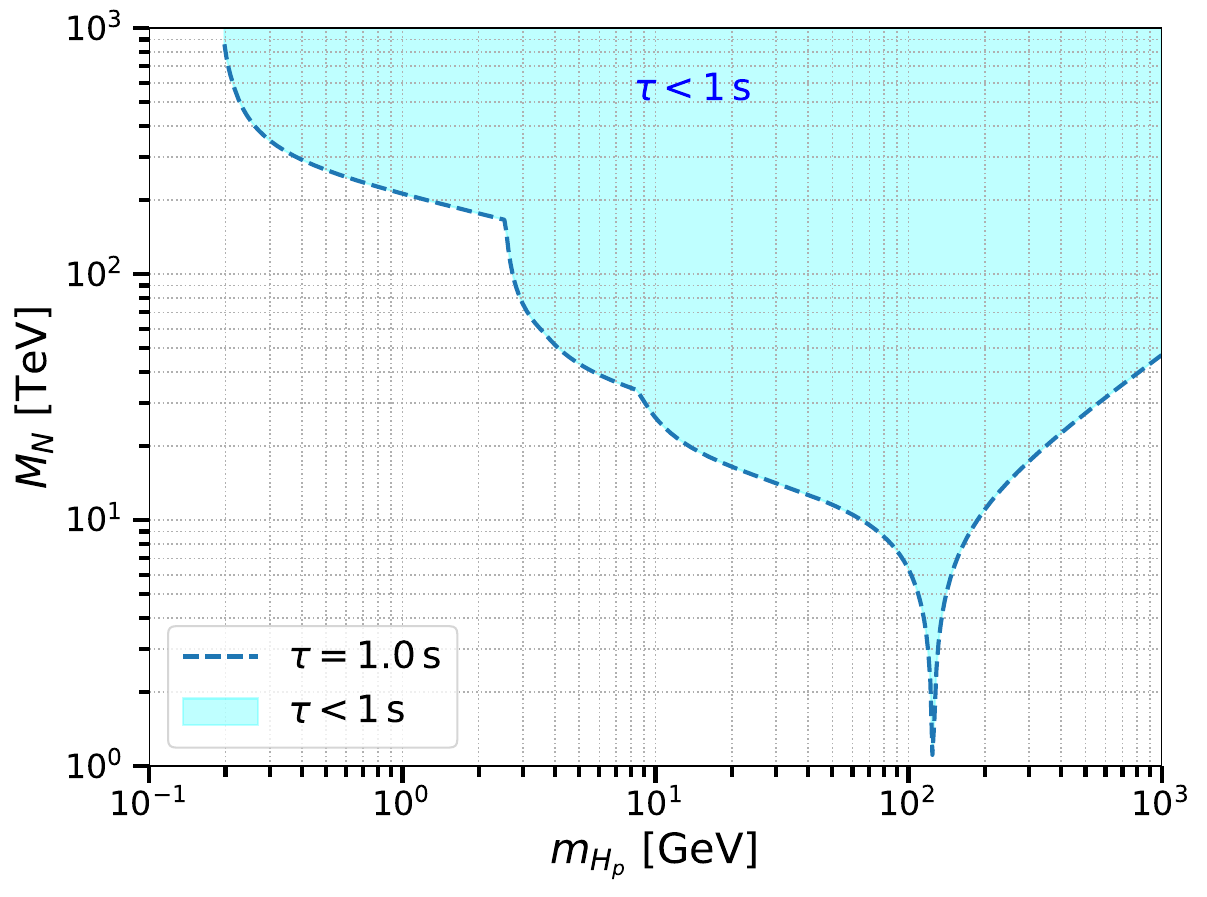}
\caption{Values of the heavy-neutrino mass $M_N$ and dark scalar mass $m_{H_p}$ for which the decay of $H_p$ occurs before BBN. We fix $y_N=1$ and $m_\nu=0.05~\text{eV}$.}
\label{fig:BBN}
\end{figure}

Figure~\ref{fig:BBN} confirms this scaling by showing the $(m_{H_p},M_N)$ region where $H_p\to f\bar f$ occurs before BBN ($\tau_{\rm ferm}<\tau_{\rm BBN}$), assuming $y_N=1$ and $m_\nu=0.05~\text{eV}$. For $m_{H_p}\lesssim 100~\text{GeV}$, larger $m_{H_p}$ requires smaller $M_N$ to keep $\tau_{\rm ferm}<1~\text{s}$. A pronounced dip appears as $m_{H_p}\to m_H$; for $m_{H_p}>m_H$, the trend reverses and the allowed $M_N$ increase roughly with $m_{H_p}$. Away from the resonance, one typically needs $M_N\gtrsim 10~\text{TeV}$ to ensure $\tau_{\rm ferm}<1~\text{s}$.

Since we work with WIMP-range $m_\chi$ and require $m_\chi>m_{H_p}$ (secluded regime), and because BBN safety favors $M_N\gtrsim 10~\text{TeV}$ to guarantee prompt $H_p$ decays into SM fermions, in what follows we assume the mass hierarchy
\begin{equation}
M_N > m_\chi > m_{H_p}\,.
\end{equation}

When several heavy Majorana eigenstates $N_I$ are present, the slowest (last) to
lose equilibrium is the \emph{lightest} one. It is therefore sufficient (and conservative)
to track a single state $N$ with mass $M_N \equiv \min_I(M_{N,I})$. In the aligned limit, we denote by $m_\nu$ the light
eigenvalue paired with this lightest $N$.

\paragraph*{Era I ($T\gg M_N$).}
For temperatures above the heavy-neutrino threshold, the SM and the hidden sector
are connected by the neutrino portal. The relevant processes are:
(i) decays/inverse decays $N \leftrightarrow L_\alpha H,\,\bar L_\alpha H^\dagger$;
(ii) $\Delta L=1$ scatterings with one neutrino Yukawa and one SM coupling,
such as $N L_\alpha \leftrightarrow Q_3 t$ (top-assisted) and
$L_\alpha A \leftrightarrow N H$ (gauge-assisted), with $A$ an electroweak gauge boson;
(iii) $\Delta L=2$ scatterings $L H \leftrightarrow \bar L H^\dagger$,
$L L \leftrightarrow H H$ induced by virtual $N$; and
(iv) portal reactions involving the dark scalar $R$ and $\chi$, including
$N R \leftrightarrow L_\alpha H$ and $R R \leftrightarrow H H$.
A complete discussion is given in App.~\ref{app:DL1-scatt}.

For the lightest state $N$ (mass $M_N$), the thermally averaged rate for the decay into $L H$ and $\bar L H^\dagger$ in the unbroken phase and aligned limit is
\begin{equation}
\langle \Gamma_{D}\rangle_T
= \Gamma_{D}\,\frac{K_1(\xi)}{K_2(\xi)},
\,\,\,\, \Gamma_D
= \frac{m_{\nu}\,M_{N}^{2}}{4\pi v_h^2}\,,
\end{equation}
with $\xi \equiv M_{N}/T$. $\langle \Gamma_{D}\rangle_T$ should be compared with $H(T)=1.66\sqrt{g_\ast}\,T^2/M_{\rm Pl}$.
Solving $\langle\Gamma_{D}\rangle_T = H(T)$ in the relativistic regime gives
\begin{equation}
\frac{T_\ast}{M_{N}}
\simeq
\left[
\frac{m_{\nu} M_{\rm Pl}}{13.28\pi\sqrt{g_\ast}\,v_h^2}
\right]^{1/3} \!\!\!\!\!
\simeq  
2.8
\left(\frac{m_{\nu}}{0.05~\mathrm{eV}}\right)^{1/3}
\left(\frac{106.75}{g_\ast}\right)^{1/6} \!\!\!\!\!,
\end{equation}
so $N$ is in thermal equilibrium with the SM bath for $T \gtrsim \text{few}\times M_N$.
At $T=M_N$ one finds
\begin{equation}
\left.\frac{\langle\Gamma_{D}\rangle_T}{H}\right|_{T=M_N}
\sim \mathcal{O}(20\text{--}30)
\quad
\text{for } m_{\nu}\sim 0.05~\mathrm{eV},
\end{equation}
showing that decays/inverse decays alone efficiently maintain contact near and
above $T\sim M_N$ for seesaw-motivated parameters.

The leading $\Delta L=1$ scatterings provide an additional contribution.
Summing the dominant top- and gauge-assisted channels one finds
%\begin{equation}
%\Gamma_{\rm scatt}^{(\Delta L=1)}(T)
%\simeq
%c_1\,(y^\dagger y)_{11}\,(y_t^2+g^2)\,T,
%\qquad
%c_1 \simeq (3\text{--}7)\times10^{-4},
%\end{equation}
%and, in the aligned limit $(y^\dagger y)_{11}=2 m_\nu M_N/v_h^2$,
\begin{equation}
\frac{\Gamma_{\rm scatt}^{(\Delta L=1)}(T)}{H(T)}
\simeq
\frac{2c_1}{1.66\sqrt{g_\ast}}\,
\frac{m_{\nu} M_{\rm Pl}}{v_h^2}(y_t^2+g^2)\,
\frac{M_{N}}{T}
\sim \mathcal{O}(1)\,\frac{M_N}{T},
\end{equation}
where $c_1 \simeq (3\text{--}7)\times10^{-4}$ and $y_t$ is the top quark Yukawa.
Thus $\Delta L=1$ scattering rates are of the same order as $H$ around $T\sim M_{N}$
(and subleading at $T\gg M_{N}$), reinforcing equilibration that is
dominated by decays/inverse decays.
The $\Delta L=2$ scatterings become efficient only at very high temperatures $T\gtrsim 10^{12}$ GeV and are negligible in the range relevant here. 

Among the portal processes, the loop-induced
$R R\leftrightarrow H H$ is negligible for our parameters, while
$N R \leftrightarrow L_\alpha H$, controlled by $y_N$ and $y_\nu$,
for $y_N=\mathcal{O}(1)$ satisfies
$\Gamma/H \sim 0.1\text{--}0.2$ near $T\sim M_{N}$. In this regime,
together with $N$ decays/inverse decays and the $\Delta L=1$ scatterings,
the $N R \leftrightarrow L_\alpha H$ channel efficiently transfers energy
between the visible and dark sectors, so that once $N$ is thermalized
the dark scalar and $\chi$ are also brought into equilibrium.

Overall, in Era~I the neutrino portal is very efficient: decays/inverse decays of $N$,
$\Delta L=1$ scatterings, and (when present with sizeable $y_N$) the
$N R\leftrightarrow L H$ channel guarantee thermal contact between the SM and the dark sector.

\begin{figure}
\includegraphics[width=0.99\linewidth]{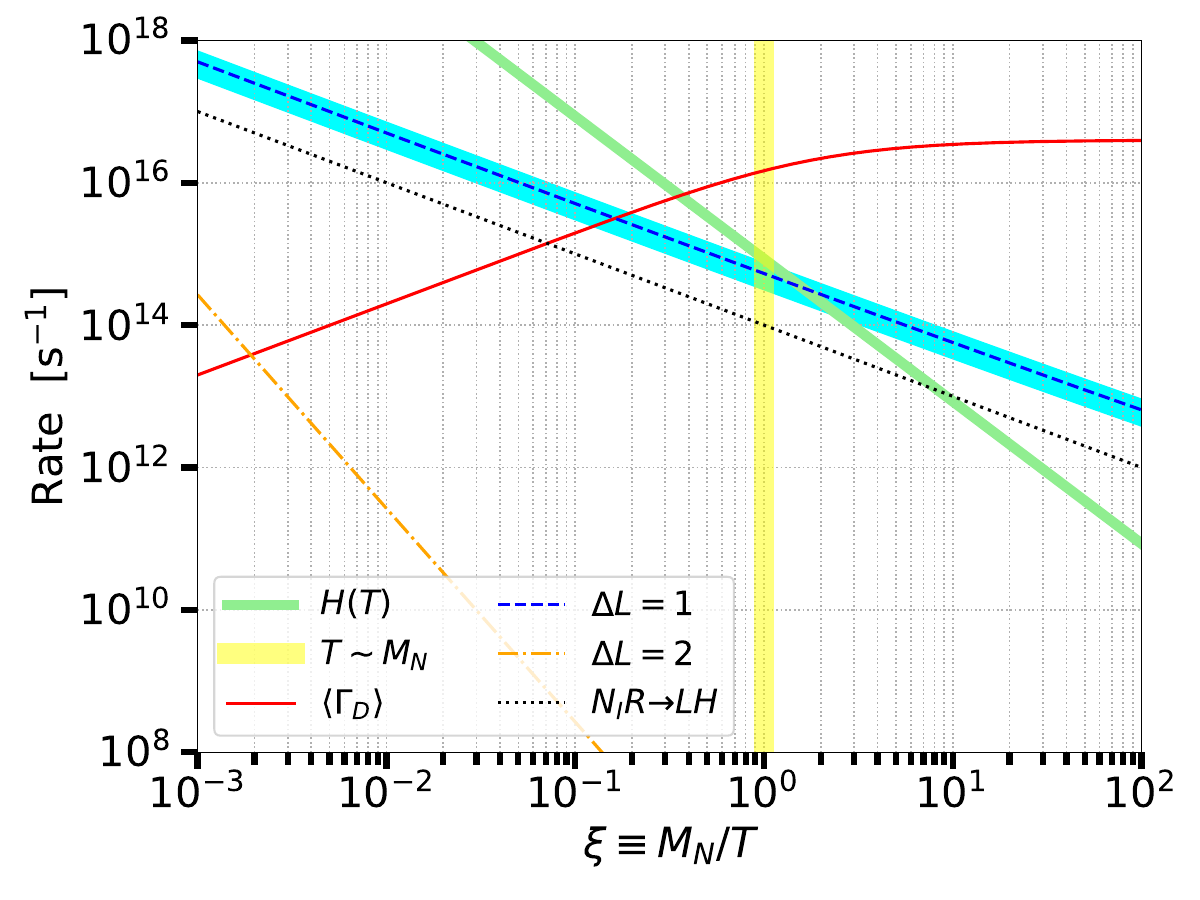}
\caption{Per-particle rates vs.\ expansion rate for the neutrino-portal processes as a function of
$\xi=M_N/T$: decay/inverse decay $N \leftrightarrow LH$ (red solid),
$\Delta L=1$ scatterings (blue dashed), $N R \leftrightarrow L_\alpha H$ (black dotted),
and $\Delta L=2$ scatterings (orange dot-dashed). We fix $M_{N}=20~\mathrm{TeV}$,
$m_\nu=0.05~\mathrm{eV}$, and $y_N=1$. The green band shows $H(T)$;
the cyan band reflects the uncertainty on the coefficient $c_1$ in the
$\Delta L=1$ rate. The vertical yellow band marks the transition to the
non-relativistic regime for $N$ where its abundance becomes Boltzmann suppressed.}
\label{fig:rate}
\end{figure}

In Fig.~\ref{fig:rate} we compare the relevant per-particle rates to $H(T)$\footnote{The figure shows \emph{per-particle} rates $\Gamma$, which by construction do not exhibit
Boltzmann suppression for $T\lesssim M_N$. That suppression enters instead through the
\emph{reaction density} $\gamma(T)\equiv n_N^{\rm eq}(T)\,\langle\Gamma\rangle$, with
$n_N^{\rm eq}(T)\propto (M_N T)^{3/2} e^{-M_N/T}$.} as a function of
$\xi$ for the benchmark $M_N=20~\mathrm{TeV}$, $m_\nu=0.05~\mathrm{eV}$,
and $y_N=1$. 
The thermally averaged decay/inverse-decay rate
$\langle\Gamma_{D}\rangle$ exceeds $H(T)$ around and above $T\sim M_N$, ensuring
efficient equilibration between $N$ and the SM bath. The $\Delta L=1$ scatterings,
shown as a band from the uncertainty in $c_1$, contribute at the level of $H$
near $T\sim M_N$ and remain comparable to $H$ for $T\gtrsim M_N$, reinforcing thermal
contact. The portal scattering $N R\leftrightarrow L_\alpha H$ is also efficient for the
chosen parameters ($y_N=\mathcal{O}(1)$), with $\Gamma/H\sim\mathcal{O}(0.1\text{--}1)$
throughout the range where $N$ is still abundant; once $N$ is in equilibrium, the dark
scalar and $\chi$ are therefore brought into equilibrium with the SM as well. In contrast,
the $\Delta L=2$ rates remain well below $H(T)$ at these scales and do not affect
equilibration. For $T \lesssim M_N$ ($\xi \gtrsim 1$), the number density of $N$ becomes
Boltzmann suppressed; all $N$-mediated processes effectively decouple, and the neutrino
portal ceases to maintain thermal contact.

\paragraph*{Era II (decoupling around $T\sim M_N$).}
As $T$ drops below the (lightest) heavy–neutrino mass, the $N$ abundance becomes Boltzmann suppressed and inter–sector energy exchange shuts off at temperature $T_{\rm{dec}}\sim (0.2-0.3) M_N$.
After visible–dark decoupling, the \emph{comoving entropies} of the two sectors are separately conserved.
Assuming the dark sector was in full chemical equilibrium at decoupling (\(\mu_i(T'_{\rm dec})=0\), where $\mu_i$ is the chemical potential for the particle $i$) and that, thereafter, both sectors remain effectively ultrarelativistic with vanishing chemical potentials \(\big(s=\tfrac{2\pi^2}{45}\,g_{*S}T^3\big)\), the temperature ratio is~\cite{Bringmann:2020mgx}
\begin{equation}
\zeta= \frac{T'}{T}=\left[\frac{g_{*S}^{\rm vis}(T)}{g_{*S}^{\rm vis}(T_{\rm dec})}\right]^{1/3}
           \left[\frac{g_{*S}^{\rm dark}(T'_{\rm dec})}{g_{*S}^{\rm dark}(T')}\right]^{1/3}.
\label{eq:xi_gstar_limit}
\end{equation}
This holds only while both baths are relativistic with \(\mu_i=0\); if any species becomes non-relativistic or \(\mu_i\!\neq\!0\), one must use the general entropy–conservation expression (see App.~\ref{app:2T-freezeout}). Whether \(\zeta=T'/T\) exceeds unity depends on the relative entropy dumps after decoupling: \(\zeta>1\) if the visible sector loses more relativistic degrees of freedom than the dark sector (and vice versa). For GeV–scale DM and multi–TeV heavy neutrinos one typically finds \(\zeta\simeq 1\text{–}2\).

\paragraph*{Era III (DM freeze–out in a hotter dark bath).\footnote{See App.~\ref{app:2T-freezeout} for the full details of the calculations reported here.}}
For $m_\chi>m_{H_p}$ (secluded regime), $\chi\chi\to H_p H_p$ keeps the dark bath thermal at $T'=\zeta T$, while the portal is far too small to re-equilibrate with the SM. Working with $x\equiv m_\chi/T$ and $x'=x/\zeta$, the Boltzmann equation for the comoving yield $Y_\chi\equiv n_\chi/s_{\rm vis}$ (defined with the \emph{visible} entropy density) reads
\begin{equation}
\frac{dY_\chi}{dx}
= - \frac{s_{\rm vis}(T)}{x\,H(T)}\,
\langle\sigma v\rangle_{\chi\chi\to H_pH_p}(T')\,
\Big[ Y_\chi^2-(Y_\chi^{\rm eq}(T'))^2 \Big],
\end{equation}
with $Y_\chi^{\rm eq}(T')=n_\chi^{\rm eq}(T')/s_{\rm vis}(T)\propto x'^2 K_2(x')$ the \emph{dark} equilibrium abundance.
During radiation domination the Hubble rate depends on the \emph{total} relativistic energy density from both sectors,
\begin{eqnarray}
\label{eq:H_general}
H(T) = \frac{\pi}{\sqrt{90}}\frac{T^{2}}{M_{\rm Pl}}
\left[g_{*}^{\rm vis}(T)+g_{*}^{\rm dark}(T')\,\zeta^{4}\right]^{1/2}, \nonumber
\end{eqnarray}
where $g_*^{\rm vis}(T)$ and $g_*^{\rm dark}(T')$ are the usual \emph{energy} degrees of freedom, and we define $g_*^{H}(T,T') = g_*^{\rm vis}(T)
+\zeta^{4}\,g^{\rm dark}_{\ast}(T')$.

The freeze–out condition is $n_\chi^{\rm eq}(T')\,\langle\sigma v\rangle(T')\ \simeq\ H(T)$ and we can use the following iterative solution to find the freeze-out temperature, which for $m_\chi$ at the GeV–TeV scale and $y_p\sim\mathcal{O}(0.1\text{–}1)$ is $x_f'\in[20,30]$, as in the standard WIMP case.
%\begin{eqnarray}
%x_f' \;\simeq\;
%\ln\!\Big(\mathcal{A}\,y_p^4\Big)
%-\frac{1}{2}\,\ln\!\Big[\ln\!\big(\mathcal{A}\,y_p^4\big)\Big], \\
%\mathcal{A}\equiv
%\frac{M_{\rm Pl}}{3.2\,\pi^{5/2} m_\chi}\,
%\frac{\xi^2}{\sqrt{g_*^{H}(T_f,T_f')}}\,,
%\end{eqnarray}
%and for $m_\chi$ at the GeV–TeV scale and $y_p\sim\mathcal{O}(0.1\text{–}1)$ one finds $x_f'\in[20,30]$, as in the standard case.
%If $g_*^{\rm dark}\xi^{4}\ll g_*^{\rm vis}$, the Hubble correction is negligible and the dominant $\xi$–dependence arises from the equilibrium density on the LHS of Eq.~\eqref{eq:RDcond}. 
The relic density then follows from the asymptotic yield $Y_{\chi,\infty}\equiv Y_\chi(T\to0)$,
\begin{equation}
\Omega_\chi h^2
\simeq
\Big(1.05\times10^9\,{\rm GeV}^{-1}\Big)\,
\frac{\sqrt{g_*^{H}(T_f,T_f')}}{g_{*S}^{\rm vis}(T_f)}
\frac{\zeta \, x_f'^2}{M_{\rm Pl}}
\frac{64\pi m_\chi^2}{3\,y_p^4}\,.
\label{eq:Omega2T-master}
\end{equation}
Inverting Eq.~\eqref{eq:Omega2T-master} yields the coupling required to reproduce the observed abundance, which is of order $\mathcal{O}(0.1-1)$,
\begin{equation}
\label{eq:yprelicp}
y_p \simeq 0.43
\left(\frac{x_f'}{25}\right)^{1/2}
\left(\frac{m_\chi}{100~\mathrm{GeV}}\right)^{1/2}
\left(\frac{g_{*S}^{\rm vis}}{86.25}\right)^{-1/4}
\left(\frac{g_*^{H}}{100}\right)^{1/8}.
\end{equation}
A convenient estimate for the required thermal cross section is $\langle\sigma v\rangle\ \approx\ 2\times 10^{-26}\ \mathrm{cm^{3}\,s^{-1}}/\zeta$
i.e.\ a \emph{hotter} bath ($\zeta>1$) requires a smaller $\langle\sigma v\rangle$ with respect to the canonical WIMP thermal cross section \cite{Bringmann:2020mgx}. %Equivalently, since $T'=\zeta T$ implies $x'=x/\zeta$ is smaller for $\zeta>1$, the larger $n_\chi^{\rm eq}(T')$ at freeze–out is compensated by a slightly reduced annihilation cross section.

\emph{Nuclear cross-section upper limits}
The upper limits from LZ \cite{LZ:2024zvo} on the spin-independent nuclear cross section $\sigma_{\rm LZ}$ provide the following constraint on the parameter space for $m_\chi=100$ GeV (see App.~\ref{app:SI-cross} for the full calculation and the results given as a function of $m_\chi$)
\begin{eqnarray}
&&\sigma^{\rm SI}_{\chi N}
\simeq
3.9\times 10^{-57}~\mathrm{cm}^2\;
\left(\frac{y_p}{1}\right)^2
\left(\frac{10~\mathrm{GeV}}{m_{H_p}}\right)^4 \times  \nonumber\\
&\times& \!\!\!
\sum_{I=1}^3
\left(\frac{y_{N,I}}{1}\right)^2
\left(\frac{m_{\nu,I}}{0.05~\mathrm{eV}}\right)^2
\left(\frac{M_{N,I}}{10\,\mathrm{TeV}}\right)^4 \lesssim \sigma_{\rm LZ}.
\label{eq:sigmaSI-numeric-3genp}
\end{eqnarray}
Fixing $y_p$ to the value that provides the correct relic density (see Eq.~\ref{eq:yprelicp}) and taking $y_N\sim 1$, Eq.~\ref{eq:sigmaSI-numeric-3genp} can be translated into an upper limit on the heavy neutrino mass. For $m_{H_p}=10~\mathrm{GeV}$ and one generation one finds
\begin{equation}
M_N
\lesssim
1.7\times 10^3~\mathrm{TeV}
\left(\frac{1}{y_p y_N}\right)^{1/2} \!\!
\left(\frac{0.05~\mathrm{eV}}{m_\nu}\right)^{1/2} \!\!
\left(\frac{m_{H_p}}{10~\mathrm{GeV}}\right).
\label{eq:MNmax-LZ}
\end{equation}

%To compensate, one needs a slightly \emph{smaller} annihilation cross section. A convenient way to express this is
%\begin{equation}
%\langle \sigma v \rangle_{\rm req}(\xi)
%\;\simeq\;
%\frac{2\times 10^{-26}\,{\rm cm}^3/{\rm s}}{\xi}\,
%\left(\frac{x_f}{20}\right)\left(\frac{\sqrt{g_\star}}{10}\right),
%\end{equation}
%so for $\xi=0.5$ one just needs roughly a factor 2 larger cross section.

\emph{Conclusions}
We introduced a minimal secluded–DM setup embedded in an extra \(U(1)_X\) with a Dirac dark fermion \(\chi\) and a Higgs–like singlet \(R\), where the Higgs–singlet portal vanishes at tree level and is \emph{radiatively} generated by the neutrino sector. The loop–induced \(R\)–\(\Phi\) mixing obeys $\tan(2\alpha) \sim \sum_I 5\times10^{-11}\, \left(\frac{m_{\nu,I}}{0.05~\mathrm{eV}}\right) \left(\frac{M_{N,I}}{10~\mathrm{TeV}}\right)^{\!2}$
linking the portal directly to the light–neutrino masses \(m_{\nu,I}\) and to the heavy singlets \(N_I\) (with masses \(M_{N,I}\)). In short, the non-observation of WIMP-like signals can be connected to the smallness of SM active neutrino masses: the tiny \(m_\nu\) makes the portal naturally minuscule, strongly suppressing direct, indirect, and collider signals today, turning the lack of detection into an \emph{natural expectation}. In particular, for heavy-neutrino masses $\gtrsim 10$ TeV, the loop–induced spin–independent nuclear cross section remains far below the sensitivity of current and foreseeable direct–detection experiments, rendering this scenario effectively invisible in standard nuclear–recoil searches. Instead, for PeV-scale heavy neutrinos the loop-induced portal is enhanced and the DM–nucleon cross section can enter the reach of direct-detection experiments. Moreover, for even larger masses $M_{N,I}\gtrsim 10^3~\mathrm{TeV}$ the model naturally enters the parametric regime of standard thermal or resonant leptogenesis, so that the out-of-equilibrium, CP-violating decays of $N_I$ can also account for the observed matter–antimatter asymmetry of the Universe \cite{Fukugita:1986hr,Buchmuller:2004nz,Davidson:2008bu}.
In the secluded regime, with $\tan(2\alpha)\ll 1$ and $m_\chi>m_{H_p}$, the relic density is set by $p$-wave annihilation $\chi\bar\chi\!\to\! H_p H_p$, and the dark-sector Yukawa couplings required to reproduce the observed abundance are $\mathcal{O}(0.1\text{--}1)$, as in the standard WIMP case.

%Despite this tiny portal, the dark matter abundance is robustly thermal. The SM and dark sectors equilibrate at \(T\gtrsim \mathrm{few}\times M_I\) via \(N_I\) decays/inverse decays and \(\Delta L=1\) scatterings, then decouple for \(T\lesssim M_I\) and evolve with different temperatures, typically with a modestly hotter dark bath (\(\xi\equiv T'/T\gtrsim 1\)). In the secluded regime \(m_\chi>m_{H_p}\), the relic density is set by the \(p\)-wave process \(\chi\bar\chi\!\to\! H_p H_p\), and values \(y_p\sim \mathcal{O}(0.1\!-\!1)\) reproduce \(\Omega_\chi h^2\simeq 0.12\). Big–Bang Nucleosynthesis is safe: with \(\tan(2\alpha)\) fixed by the loop relation, \(H_p\) decays before \(t\sim1~\mathrm{s}\) for multi–TeV \(M_I\) (or lighter \(H_p\)).

\begin{acknowledgments}
M.D.M. thanks N.Fornengo, S.Gariazzo for reading the paper and providing us very helpful comments and suggestions.
M.D.M. acknowledges support from the research grant {\sl TAsP (Theoretical Astroparticle Physics)} funded by Istituto Nazionale di Fisica Nucleare (INFN) and from the Italian Ministry of University and Research (MUR), PRIN 2022 ``EXSKALIBUR – Euclid-Cross-SKA: Likelihood Inference Building for Universe’s Research'', Grant No. 20222BBYB9, CUP I53D23000610 0006, and from the European Union -- Next Generation EU.
\end{acknowledgments}

\newpage

\onecolumngrid
\appendix

\newpage

\section{General Details of the Model}
\label{sec:model}

In this appendix we provide the full model details. Section~\ref{sec:lagrmodel} presents the Lagrangian, particle content, and the most relevant couplings. In Sec.~\ref{sec:RPHI} we describe the $\Phi$–$R$ scalar mixing. In Sec.~\ref{sec:interactions} we collect the interaction terms in the mass basis. In Sec.~\ref{sec:XSGamma} we list useful decay rates and annihilation cross sections.

\subsection{Lagrangian and Particle Content}
\label{sec:lagrmodel}

We consider a minimal UV–complete BSM dark sector that realizes the secluded mechanism \cite{Pospelov:2007mp,Pospelov:2008jd,DiMauro:2025jsb}. The model features an extra abelian gauge group $U(1)_X$ containing a Dirac fermion $\chi$ with charge $q_X^\chi$ (singlet under the SM gauge groups), a massive gauge boson $Z'$ with coupling $g_X$, three right-handed neutrinos $N_i$ ($i=1,\ldots,3$), and a complex scalar singlet $R$ that mixes with the SM Higgs doublet $\Phi$ (see e.g.~\cite{Baek:2011aa,Baek:2013qwa}). The relevant Lagrangian terms are given by
\begin{eqnarray}
\mathcal{L} &\supset&
-\frac{1}{4} F'_{\mu\nu}F'^{\mu\nu}
-\frac{\epsilon}{2}\, F'_{\mu\nu} B^{\mu\nu}
+ \partial_\mu R^\dagger \partial^\mu R
- V(\Phi,R)
+ \bar\chi\,(i\!\not\!\! D - m_\chi)\,\chi
\nonumber\\[2pt]
&&
-\Big( y_p\,\bar\chi\chi\,R + \text{h.c.}\Big)
-\Big(Y_\nu^{\alpha i}\,\bar L_\alpha\tilde{\Phi} N_i
+ \tfrac12 Y_N^{ij}\,R\,\overline{N_i^{\;c}}N_j + \text{h.c.}\Big),
\label{eq:Ltot}
\end{eqnarray}
where $L_\alpha=(\nu_{\alpha L},\ell_{\alpha L})^T$ are the SM lepton doublets, $\Phi$ is the SM Higgs, $B_{\mu\nu}$ is the hypercharge field–strength tensor, and $Y_\nu$ ($3\times 3$ complex) and $Y_N=Y_N^T$ (complex symmetric) are the neutrino Yukawa matrices.

In what follows we take $\chi$ to be vector-like, $q_{\chi_L}=q_{\chi_R}=q_\chi$. Then a bare mass $m_0 \bar\chi\chi$ is gauge invariant if $R$ is neutral ($q_X^R=0$), and we include the renormalizable Yukawa portal $-\,y_p \bar\chi\chi R$. In this realization $U(1)_X$ is unbroken by $R$ acquiring a VEV, so the $Z'$ mass can be generated through a Stückelberg mechanism.
The Yukawa interaction $y_p \bar\chi\chi R$ controls the coupling of DM to the dark scalar, and since $\chi$ is vector-like,
\begin{equation}
m_\chi \;=\; m_0 + \frac{y_p v_r}{\sqrt{2}}\,.
\end{equation}
Since $R$ is neutral ($q_X^R=0$), gauge invariance of the $R\overline{N^c}N$ term implies that the heavy neutrinos $N_i$ are also neutral under $U(1)_X$. Therefore, $N_I$ do not couple to $Z'$.

The covariant derivative is $D_\mu=\partial_\mu + i\,q_X g_X Z'_\mu$ (e.g.\ $q_X\!\to\! q_X^\chi$ on $\chi$). We define $\alpha_X \equiv g_X^2/(4\pi)$. Here $F'_{\mu\nu}$ and $g_X$ denote the field strength and coupling of $Z'_\mu$. The term $-\frac{\epsilon}{2}\, F'_{\mu\nu} B^{\mu\nu}$ represents kinetic mixing between $U(1)_X$ and hypercharge, controlled by $\epsilon$~\cite{Holdom:1985ag,Essig:2013lka}. SM fermions are neutral under $U(1)_X$, so they couple to $Z'$ only via kinetic mixing at tree level.

\subsection{$R$–$\Phi$ Mixing}
\label{sec:RPHI}

The scalar potential is
\begin{eqnarray}
V(\Phi,R)
&=& \mu_H^2\,\Phi^\dagger\Phi + \mu_R^2\, R^\dagger R
+\lambda_H (\Phi^\dagger\Phi)^2
+\lambda_R (R^\dagger R)^2
+\kappa (\Phi^\dagger\Phi)(R^\dagger R)\, .
\end{eqnarray}
After EWSB,
\begin{equation}
\Phi=\begin{pmatrix} G^+ \\  \dfrac{v_h+h+i G^0}{\sqrt{2}} \end{pmatrix}, \,\,
\tilde{\Phi} = i \sigma_2 \Phi^\ast =
\begin{pmatrix}
\dfrac{v_h + h - i\,G^0}{\sqrt{2}} \\
-\,G^-
\end{pmatrix},
\,\,
R=\dfrac{v_r+\rho}{\sqrt{2}}\,,
\end{equation}
with $v_h\simeq 246~\text{GeV}$ the SM Higgs vacuum expectation value (VEV) and $v_r$ the singlet VEV; $G^+$ and $G^0$ are eaten by $W^\pm$ and $Z$. Since $U(1)_X$ is unbroken by $R$, no extra Goldstone appears. In the CP-even basis $(h,\rho)$,
\begin{equation}
\mathcal{M}^2=
\begin{pmatrix}
2\lambda_H v_h^2 & \kappa v_h v_r \\
\kappa v_h v_r & 2\lambda_R v_r^2
\end{pmatrix},
\qquad
\tan 2\alpha = \frac{\kappa \, v_h v_r}{\lambda_R v_r^2 - \lambda_H v_h^2}\,,
\label{eq:mixing}
\end{equation}
and the mass eigenstates are
\begin{equation}
\begin{pmatrix} H \\ H_p \end{pmatrix}
=
\begin{pmatrix}
\cos\alpha & \ \ \sin\alpha \\
-\sin\alpha & \ \ \cos\alpha
\end{pmatrix}
\begin{pmatrix} h \\ \rho \end{pmatrix},
\quad\Rightarrow\quad
h=\cos\alpha\,H-\sin\alpha\,H_p,\qquad
\rho=\sin\alpha\,H+\cos\alpha\,H_p,
\label{eq:mixing-HP}
\end{equation}
with masses
\begin{equation}
m_{H,H_p}^2=\lambda_H v_h^2 + \lambda_R v_r^2 \mp
\sqrt{(\lambda_R v_r^2-\lambda_H v_h^2)^2+(\kappa v_h v_r)^2}.
\end{equation}

\subsection{Interaction Terms in the Mass Basis}
\label{sec:interactions}

We now expand the interaction terms for $H,H_p$ with SM and dark fermions.

\paragraph{(i) SM fermions.}
The SM Yukawa Lagrangian
$-\mathcal{L}\supset y_f\,\bar f_L \Phi f_R + \text{h.c.}$ with $m_f=y_f v_h/\sqrt{2}$ gives, after EWSB and Eq.~\eqref{eq:mixing-HP},
\begin{equation}
\mathcal{L}_{H,H_p\!-\!f\bar f}
\;=\;
-\,\sum_f \frac{m_f}{v_h}\,
\big(\cos\alpha\,H-\sin\alpha\,H_p\big)\;\bar f f\,,
\label{eq:LffH}
\end{equation}
so
$g_{H f\bar f} = \frac{m_f}{v_h}\cos\alpha$ and
$g_{H_p f\bar f} = -\,\frac{m_f}{v_h}\sin\alpha$.

\paragraph{(ii) Dark matter.}
From $-\mathcal{L}\supset y_p\,\bar\chi\chi\,R$,
\begin{equation}
\mathcal{L}_{H,H_p\!-\!\chi\bar\chi}
\;=\;
-\,\frac{y_p}{\sqrt{2}}\,
\big(\sin\alpha\,H+\cos\alpha\,H_p\big)\;\bar\chi\chi\,,
\label{eq:LchiH}
\end{equation}
so
$g_{H\chi\bar\chi}=\frac{y_p}{\sqrt{2}}\sin\alpha$ and
$g_{H_p\chi\bar\chi}=\frac{y_p}{\sqrt{2}}\cos\alpha$.

\paragraph{(iii) Heavy (right-handed) neutrinos.}
From $-\mathcal{L}\supset \tfrac12\,Y_N^{ij}\,R\,\overline{N_i^{\,c}}N_j+\text{h.c.}$ one has $M_N=\tfrac{v_r}{\sqrt{2}}\,Y_N$ and
\begin{equation}
\mathcal{L}_{\rho NN}
\;=\;
-\,\frac{1}{2\sqrt{2}}\;\rho\;\overline{N^{\,c}}\,Y_N\,N+\text{h.c.}
\end{equation}
In the heavy-neutrino mass basis, $U_N^T M_N U_N=\widehat M_N=\mathrm{diag}(M_1,M_2,M_3)$ and $U_N^T Y_N U_N = \sqrt{2}\,\widehat M_N/v_r$, so
\begin{equation}
\mathcal{L}_{H,H_p\!-\!N N}
\;=\;
-\,\sum_{I=1}^3 \frac{M_I}{2 v_r}\;
\big(\sin\alpha\,H+\cos\alpha\,H_p\big)\;
\overline{N_I^{\,c}}\,N_I\;+\;\text{h.c.}
\label{eq:LNNH}
\end{equation}
Equivalently, in terms of $y_{N,I}$ with $M_I=y_{N,I}v_r/\sqrt{2}$,
$\mathcal{L}\supset -\sum_I \tfrac{1}{2\sqrt{2}}\,y_{N,I}\,(\sin\alpha\,H+\cos\alpha\,H_p)\,\overline{N_I^{\,c}}N_I+\text{h.c.}$
Therefore $g_{N_I N_I H_p}  = \frac{y_{N,I}}{\sqrt{2}}\,\cos\alpha$ and $g_{N_I N_I H}   = -\,\frac{m_f}{v_h}\,\sin\alpha$.

\paragraph{(iv) Light–heavy neutrino Yukawas (Dirac sector).}
From $-\mathcal{L}\supset Y_\nu^{\alpha i}\,\bar L_\alpha\,\tilde\Phi\,N_i+\text{h.c.}$ one obtains $m_D=Y_\nu v_h/\sqrt{2}$ and
\begin{equation}
\mathcal{L}_{H,H_p\!-\!\nu N}
\;=\;
-\,\frac{1}{\sqrt{2}}\,
\big(\cos\alpha\,H-\sin\alpha\,H_p\big)\;
\bar\nu_{L\alpha}\,Y_\nu^{\alpha i}\,N_i
\;+\;\text{h.c.},
\label{eq:LnuNH}
\end{equation}
which directly gives the $H,H_p$ couplings to light–heavy pairs in the mass basis (see also Sec.~\ref{sec:ibarra} for the aligned limit).
The light–heavy neutrino couplings are encoded in $Y_\nu$ as they appear in Eq.~\eqref{eq:LnuNH}; they are generally non-diagonal and become diagonal only in the aligned limit (Sec.~\ref{sec:ibarra}).

\paragraph{(v) Scalar potential (for reference).}
From
$V(\Phi,R)=\mu_H^2\,\Phi^\dagger\Phi+\mu_R^2 R^\dagger R+\lambda_H(\Phi^\dagger\Phi)^2+\lambda_R(R^\dagger R)^2+\kappa(\Phi^\dagger\Phi)(R^\dagger R)$,
the cubic interactions in the gauge basis are
\begin{equation}
V_{\rm cubic}\supset
\lambda_H v_h\,h^3
+\lambda_R v_r\,\rho^3
+\frac{\kappa}{2}\,v_h\,h\,\rho^2
+\frac{\kappa}{2}\,v_r\,h^2\rho\,.
\end{equation}
Using Eq.~\eqref{eq:mixing-HP} one can obtain all $H,H_p$ trilinears (e.g.\ $HHH$, $HHH_p$, $HH_pH_p$, $H_pH_pH_p$). These are relevant for scalar cascades and self-scattering but not needed for the fermionic interactions above.

\subsection{Useful Decay Rates and Annihilation Cross Sections}
\label{sec:XSGamma}

\subsubsection{Decay rate of $H_p$}
\label{sec:Hp-decays}

The couplings in Sec.~\ref{sec:interactions} fix the tree-level partial widths of $H_p$ into fermion pairs. For a Dirac fermion $f$ with $\mathcal{L}\supset -g_{H_p f\bar f} H_p \bar f f$,
\begin{equation}
\Gamma(H_p \to f\bar f)
=
N_c^f \,\frac{g_{H_p f\bar f}^2}{8\pi}\,m_{H_p}
\left(1-\frac{4m_f^2}{m_{H_p}^2}\right)^{3/2},
\end{equation}
where $N_c^f$ is the color factor. Using $g_{H_p f\bar f} = -(m_f/v_h)\sin\alpha$,
\begin{equation}
\Gamma_{\rm ferm}(H_p)
=
\sum_{f}
\frac{N_c^f\,m_{H_p}}{8\pi}\,
\frac{m_f^2}{v_h^2}\,\sin^2\!\alpha\,
\left(1-\frac{4m_f^2}{m_{H_p}^2}\right)^{3/2},
\label{eq:Hp_SMfermions}
\end{equation}
summing over all kinematically allowed $f$ with $m_{H_p}>2m_f$.

For the DM channel, $g_{H_p \chi\bar\chi} = \tfrac{y_p}{\sqrt{2}}\cos\alpha$ gives (for $m_{H_p}>2m_\chi$)
\begin{equation}
\Gamma(H_p \to \chi\bar\chi)
=
\frac{y_p^2 \cos^2\!\alpha}{16\pi}\,m_{H_p}
\left(1-\frac{4m_\chi^2}{m_{H_p}^2}\right)^{3/2}.
\label{eq:Hp_DM}
\end{equation}
If open, heavy-neutrino final states $H_p\to N_I N_I$ contribute with $g_{H_p N_I N_I} = \tfrac{y_{N,I}}{\sqrt{2}}\cos\alpha$ (Majorana prefactor understood). In the parameter region of interest, the branching ratios are controlled by the competition between the mixing-suppressed SM channels $\propto \sin^2\!\alpha$ and the direct dark-portal channel $\propto y_p^2\cos^2\!\alpha$. For small $\alpha$ and $m_{H_p}>2m_\chi$, one typically has ${\rm BR}(H_p\to\chi\bar\chi)\simeq 1$, unless sizable $H_p\to N_I N_I$ is kinematically allowed.

\subsubsection{Tree-level \texorpdfstring{$\chi\bar\chi\to H_p H_p$}{χχ→HpHp} and the \texorpdfstring{$p$}{p}-wave coefficient}
\label{app:xschichiHpHp}

We take a Dirac fermion \(\chi\) of mass \(m_\chi\) coupled to a real scalar \(H_p\) of mass \(m_{H_p}\) via
\begin{equation}
\mathcal{L}\supset y_p\,\bar\chi\chi\,H_p.
\end{equation}
The thermal average cross section comes from the $t/u$ diagrams, it is $p$-wave and given by the following expression
\begin{equation}
\sigma v_{\rm rel}\big(\chi\bar\chi\to H_p H_p\big)
=
\frac{y_p^4}{64\pi m_\chi^2}\;
\frac{\sqrt{1-r}}{\big(1-\tfrac{r}{2}\big)^{4}}
\left(1-r+\frac{r^{2}}{8}\right)\;
v_{\rm rel}^2
\;+\;\mathcal{O}(v_{\rm rel}^4)\,,
\end{equation}
where $r=m_{H_p}^2/m_\chi^2$.

\section{Light and heavy neutrino masses, interactions, and the seesaw mechanism}
\label{sec:neutrino}

In this appendix we explain the setup for the light and heavy neutrino Yukawa
couplings, masses, and the type-I seesaw mechanism. In
Sec.~\ref{sec:neutrinos} we discuss the Yukawa interactions of light and heavy
neutrinos with the scalars; in Sec.~\ref{sec:seesaw} we report the main steps
of the type-I seesaw mechanism; and in Sec.~\ref{sec:intmass} we list the
couplings in the mass eigenstate basis. The details of the neutrino Yukawa
interactions and the seesaw mechanism are inspired by
Refs.~\cite{Casas:2001sr,King:2003jb,Davidson:2008bu,Heeck:2012fw,Cordero-Carrion:2019qvb}.

\subsection{Light and heavy neutrino Yukawa interactions}
\label{sec:neutrinos}

The interactions between light and heavy neutrinos and the scalars $\Phi$ and
$R$ are given in Eq.~\eqref{eq:Ltot}:
\begin{equation}
\mathcal{L}_Y \supset
-\;Y_\nu^{\alpha i}\,\overline{L_\alpha}\,\tilde{\Phi}\,N_i
-\frac{1}{2}\,Y_N^{ij}\,R\,\overline{N_i^{\;c}} N_j
+\text{h.c.}\,,
\label{eq:LY-flavour}
\end{equation}
where $L_\alpha$ are the left-handed lepton doublets (with
$\alpha=e,\mu,\tau$ labelling lepton flavours), $i,j=1,2,3$ label right-handed
singlet neutrinos, $Y_\nu$ is a generic complex $3\times 3$ matrix, and
$Y_N = Y_N^T$ is complex symmetric.

\emph{Dirac term.}
After EWSB the first term in Eq.~\eqref{eq:LY-flavour} can be written as
\begin{align}
-\;Y_\nu^{\alpha i}\,\overline{L_\alpha}\,\tilde{\Phi}\,N_i &=
-\,Y_\nu^{\alpha i}\!
\left[
\overline{\nu_{\alpha L}}\,
\frac{v_h + h - iG^0}{\sqrt{2}}
\;-\;
\overline{\ell_{\alpha L}}\,G^-
\right] N_i .
\end{align}
Focusing on the neutral part and switching to matrix notation,
\begin{equation}
-\,\overline{\nu_L}\,
\frac{v_h + h - iG^0}{\sqrt{2}}\,
Y_\nu\,N
+\text{h.c.} = 
-\,\overline{\nu_L}\,
\underbrace{\Big(\frac{v_h}{\sqrt{2}}Y_\nu\Big)}_{\displaystyle m_D}
N
-\frac{1}{\sqrt{2}}\,
\overline{\nu_L}\,h\,Y_\nu\,N
+\frac{i}{\sqrt{2}}\,
\overline{\nu_L}\,G^0\,Y_\nu\,N
+\text{h.c.}
\label{eq:Dirac-mass-def}
\end{equation}
We identify in the Dirac mass term $-\overline{\nu_L}\, m_D N$ the
$3\times 3$ Dirac mass matrix $m_D \equiv v_h Y_\nu/\sqrt{2}$. The
remaining terms in Eq.~\eqref{eq:Dirac-mass-def} describe the couplings
of $\nu_L$ to the physical Higgs field $h$ and to the neutral Goldstone
$G^0$.

\medskip

\emph{Majorana term.}
For the singlet sector, one has
\begin{align}
-\frac{1}{2}\,Y_N^{ij}\,R\,\overline{N_i^{\;c}} N_j
&=
-\frac{1}{2}\,\frac{v_r + \rho}{\sqrt{2}}\,
Y_N^{ij}\,\overline{N_i^{\;c}} N_j 
=
-\frac{1}{2}\,\overline{N^{c}}\,
\underbrace{\Big(\frac{v_r}{\sqrt{2}}Y_N\Big)}_{\displaystyle M_N}
N
-\frac{1}{2\sqrt{2}}\,
\overline{N^{c}}\,
\rho\,Y_N\,N.
\end{align}
Thus from the Majorana mass term
$-\tfrac{1}{2}\,\overline{N^{c}} M_N N$,
we identify the heavy-neutrino Majorana mass matrix
$M_N \equiv v_r Y_N/\sqrt{2}$, which is complex symmetric,
and the coupling of $\rho$ to $N_i N_j$ is proportional to $Y_N$.

\subsection{Type-I seesaw mechanism}
\label{sec:seesaw}

Collecting the VEV-induced pieces, the neutral-fermion mass Lagrangian is
\begin{equation}
\mathcal{L}_{\rm mass}
\supset
-\,\overline{\nu_L}\,m_D\,N
-\frac{1}{2}\,\overline{N^{c}}\,M_N\,N
+\text{h.c.}
\label{eq:L-mass-2block}
\end{equation}
It is convenient to write this in a symmetric $6\times 6$ form.
Define
\begin{equation}
n_L \equiv
\begin{pmatrix}
\nu_L \\
N^{c}
\end{pmatrix},
\qquad
n_L^c =
\begin{pmatrix}
\nu_L^{c} \\
N
\end{pmatrix}.
\end{equation}
Using
$\overline{\nu_L} m_D N
=
\frac{1}{2}\big(\overline{\nu_L} m_D N
+ \overline{N^{c}} m_D^T \nu_L^{c}\big)$
and similarly for the Majorana term, we obtain
\begin{equation}
\mathcal{L}_{\rm mass}
\supset
-\frac{1}{2}\;
\overline{n_L^{\,c}}\,
\mathcal{M}\,
n_L
+\text{h.c.},
\qquad
\mathcal{M} =
\begin{pmatrix}
0      & m_D \\
m_D^T  & M_N
\end{pmatrix},
\label{eq:full-6x6-mass}
\end{equation}
where $\mathcal{M}$ is complex symmetric, as appropriate for Majorana masses.

\medskip

\emph{Type-I seesaw: matrix diagonalization.}
In the seesaw regime $\|m_D M_N^{-1}\|\ll 1$, we define
\begin{equation}
\Theta \equiv m_D M_N^{-1},
\end{equation}
and the block-antihermitian generator
\begin{equation}
\Omega =
\begin{pmatrix}
0           & \Theta \\
-\Theta^\dagger & 0
\end{pmatrix},
\end{equation}
which can be used to approximately block-diagonalize $\mathcal{M}$ via
an (approximately) unitary transformation $U_I \simeq \exp(-\Omega)$.
To second order in $\Theta$ one finds
\begin{equation}
U_I^T \mathcal{M} U_I =
\begin{pmatrix}
M_\nu & 0\\[2pt]
0 & M_{\rm heavy}
\end{pmatrix}
+ \mathcal{O}(\Theta^3),
\end{equation}
with
\begin{align}
M_\nu
&= -\,m_D M_N^{-1} m_D^T
= -\,\Theta M_N \Theta^T
+ \mathcal{O}(\Theta^4 M_N),
\label{eq:ss-mnu-typeI}
\\[4pt]
M_{\rm heavy}
&= M_N + \frac{1}{2}\!\left(\Theta^\dagger m_D + m_D^T\Theta\right)
+ \mathcal{O}(\Theta^4 M_N).
\label{eq:ss-Mheavy}
\end{align}
Thus $M_{\rm heavy}\simeq M_N$ up to relative $\mathcal{O}(\Theta^2)$ corrections.

In the one-generation case with real parameters $m_D$ and $M_N$, the eigenvalues are
\begin{equation}
m_{\rm light}\simeq -\,\frac{m_D^2}{M_N},
\qquad
m_{\rm heavy}\simeq M_N\left(1+\frac{m_D^2}{M_N^2}\right).
\end{equation}
In terms of the Yukawa couplings, using
$m_D = y_\nu v_h/\sqrt{2}$ and $M_N = y_N v_r/\sqrt{2}$, the seesaw relation
for one generation can be written as
\begin{equation}
m_\nu \;\simeq\; -\,\frac{m_D^2}{M_N}
= -\,\frac{y_\nu^2 v_h^2}{\sqrt{2}\,y_N v_r}\,,
\qquad
M_{\rm heavy}
\simeq M_N
= \frac{y_N v_r}{\sqrt{2}}.
\label{eq:seesaw-1gen}
\end{equation}

\subsection{Light and heavy neutrino mass eigenstates and Yukawa interactions in this basis}
\label{sec:intmass}

\emph{Heavy (RH) neutrino mass eigenstates.}
Since $M_N$ is complex symmetric, it admits a Takagi factorization: there exists
a unitary matrix $U_N$ such that
\begin{equation}
U_N^T M_N U_N
\;=\;
\widehat M_N
\;\equiv\;
\mathrm{diag}(M_1,M_2,M_3),
\qquad M_I > 0.
\label{eq:MN-diag}
\end{equation}
We define the heavy-neutrino mass eigenstates $N_I$ via
\begin{equation}
N_i = (U_N)_{iI}\,N_I,
\label{eq:N-rotation}
\end{equation}
so that the mass term becomes
\begin{equation}
\mathcal{L}_M
\supset -\frac{1}{2}\, \overline{N^{c}}\,M_N\,N
= -\frac{1}{2}\, \overline{N^{c}}_I\,(U_N^T M_N U_N)_{IJ} N_J
= -\frac{1}{2}\sum_{I} M_I\,\overline{N_I^{\,c}} N_I\,.
\end{equation}
Equivalently, in terms of the diagonal Yukawas,
\begin{equation}
\widehat Y_N
\;\equiv\;
U_N^T Y_N U_N
=
\mathrm{diag}(y_{1,N},y_{2,N},y_{3,N}),
\qquad
M_I = \frac{v_r}{\sqrt{2}}\,y_{I,N}\,.
\label{eq:YN-diag}
\end{equation}
In this basis ($\widehat M_N$ diagonal), the Dirac mass matrix becomes
\begin{equation}
m_D \;\to\; m_D' = m_D\,U_N
= \frac{v_h}{\sqrt{2}}\,Y_\nu',
\qquad
Y_\nu' \equiv Y_\nu U_N,
\label{eq:mD-prime}
\end{equation}
which is generically non-diagonal.

\medskip

\emph{Light neutrinos and type-I seesaw.}
In the basis with diagonal $\widehat M_N$, the seesaw formula
\eqref{eq:ss-mnu-typeI} reads
\begin{equation}
M_\nu
\simeq -\,m_D M_N^{-1} m_D^T
= -\,m_D' \,\widehat M_N^{-1}\,m_D'^T
\;=\;
-\,\frac{v_h^2}{2}\,Y_\nu'\,\widehat M_N^{-1}\,Y_\nu'^T.
\label{eq:seesaw-Mnu}
\end{equation}
The light-neutrino mass matrix $M_\nu$ is complex symmetric and is diagonalized
by a unitary matrix $U_\nu$:
\begin{equation}
U_\nu^T\,M_\nu\,U_\nu
\;=\;
\widehat m_\nu
\;\equiv\;
\mathrm{diag}(m_1,m_2,m_3),
\qquad m_i \ge 0.
\label{eq:Mnu-diag}
\end{equation}
The light-neutrino mass eigenstates $\nu_i$ are defined by
\begin{equation}
\nu_{L\alpha}
=
(U_\nu)_{\alpha i}\,\nu_{iL},
\label{eq:nu-rotation}
\end{equation}
up to rephasings; $U_\nu$ can be identified with the PMNS matrix in the usual
way.

\medskip

\emph{Yukawa interactions in the mass basis.}
From Eq.~\eqref{eq:Dirac-mass-def} and the $(h,\rho)$--$(H,H_p)$ mixing
in Eq.~\eqref{eq:mixing-HP}, the couplings of the CP-even scalars to
$\nu$ and $N$ in the flavour basis are
\begin{equation}
\mathcal{L}_{H,H_p\!-\!\nu N}
=
-\frac{1}{\sqrt{2}}\,
\big(\cos\alpha\,H-\sin\alpha\,H_p\big)\;
\overline{\nu_{L\alpha}}\,
Y_\nu^{\alpha i}\,
N_i
+\text{h.c.}
\label{eq:L-H-nuN-flavour}
\end{equation}
We now rotate to the light and heavy mass bases using
Eqs.~\eqref{eq:N-rotation} and \eqref{eq:nu-rotation}, obtaining
\begin{equation}
\mathcal{L}_{H,H_p\!-\!\nu N}
=
-\frac{1}{\sqrt{2}}\,
\big(\cos\alpha\,H-\sin\alpha\,H_p\big)\;
\overline{\nu_{iL}}\,
\big(Y_\nu^{\rm mass}\big)_{iI}\,
N_I
+\text{h.c.},
\label{eq:L-H-nuN-mass}
\end{equation}
where
\begin{equation}
Y_\nu^{\rm mass}
\;\equiv\;
U_\nu^\dagger\,Y_\nu\,U_N.
\label{eq:Ynu-mass-def}
\end{equation}
In general, $Y_\nu^{\rm mass}$ is \emph{not} diagonal. The couplings
\begin{equation}
g_{H\,\nu_i N_I}
= -\,\frac{\cos\alpha}{\sqrt{2}}\,(Y_\nu^{\rm mass})_{iI},
\qquad
g_{H_p\,\nu_i N_I}
= +\,\frac{\sin\alpha}{\sqrt{2}}\,(Y_\nu^{\rm mass})_{iI},
\label{eq:g-H-couplings}
\end{equation}
show that each heavy neutrino $N_I$ couples to all light eigenstates $\nu_i$.

\subsection{Casas--Ibarra parametrization and aligned limit}
\label{sec:ibarra}

Given the low-energy neutrino data (i.e.~the light eigenvalues $m_i$ and the
PMNS matrix $U_\nu$, including CP phases) and a choice of heavy-neutrino masses
$M_I$ (the eigenvalues of $M_N$), the most general Dirac mass matrix $m_D$
reproducing the type-I seesaw relation
\begin{equation}
M_\nu \;=\; -\,m_D M_N^{-1} m_D^T
\end{equation}
can be written in the Casas--Ibarra form
\begin{equation}
m_D
=
i\,U_\nu\,\sqrt{\widehat m_\nu}\,O\,\sqrt{\widehat M_N},
\qquad
O^T O = \mathbb{I},
\label{eq:ss-CI}
\end{equation}
where
\begin{equation}
\widehat m_\nu = \mathrm{diag}(m_1,m_2,m_3),
\qquad
\widehat M_N = \mathrm{diag}(M_1,M_2,M_3),
\end{equation}
and $O$ is a complex orthogonal matrix encoding the residual high-energy
freedom not fixed by low-energy observables. 

\medskip

In general $O$ is not diagonal, so Eq.~\eqref{eq:Ynu-mass-def} shows that
$Y_\nu^{\rm mass}$ has off-diagonal entries: each heavy state $N_I$ couples to
\emph{all} light mass eigenstates $\nu_i$. Consequently, the seesaw relation
\(
M_\nu = - m_D M_N^{-1} m_D^T
\)
involves coherent sums over $I$, and one cannot identify a light mass $m_i$
with a single Yukawa coupling via a simple $m_i = y_{i,\nu} v_h/\sqrt{2}$.

A special simplification occurs only in an \emph{aligned limit}. Work in the
basis where the heavy-neutrino mass matrix is diagonal,
$M_N = \widehat M_N = \mathrm{diag}(M_1,M_2,M_3)$, and write the Dirac Yukawas
in the light- and heavy-mass bases as
\begin{equation}
Y_\nu^{\rm mass}
= i\,\frac{\sqrt{2}}{v_h}\,
\sqrt{\widehat m_\nu}\,O\,\sqrt{\widehat M_N},
\label{eq:Ymass-CI-again}
\end{equation}
with $\widehat m_\nu = \mathrm{diag}(m_1,m_2,m_3)$ and $O$ a complex orthogonal
matrix. For generic $O$ the matrix $Y_\nu^{\rm mass}$ is not diagonal, so each
light eigenstate $\nu_i$ couples to several heavy states $N_I$ and its mass
$m_i$ arises from a coherent combination of $(Y_\nu^{\rm mass})_{iI}$ entries.
In this generic situation it is \emph{not} correct to write
$m_i = y_{i,\nu} v_h/\sqrt{2}$ with a single Yukawa $y_{i,\nu}$.

In contrast, in the aligned limit we choose
\begin{equation}
O = \mathbb{1},
\end{equation}
and use the freedom to relabel the heavy mass eigenstates so that the indices
of $\widehat m_\nu$ and $\widehat M_N$ are paired consistently. Then
Eq.~\eqref{eq:Ymass-CI-again} reduces to
\begin{equation}
Y_\nu^{\rm mass}
= i\,\frac{\sqrt{2}}{v_h}\,
\sqrt{\widehat m_\nu}\,\sqrt{\widehat M_N}
= i\,\mathrm{diag}\!\left(
\frac{\sqrt{2 m_1 M_1}}{v_h},
\frac{\sqrt{2 m_2 M_2}}{v_h},
\frac{\sqrt{2 m_3 M_3}}{v_h}
\right).
\end{equation}
Therefore, in the aligned limit ($O=\mathbb{1}$ with a consistent pairing of
$m_{\nu,I}$ and $M_I$), the Yukawa matrix in the mass basis is diagonal and one
finds
\begin{equation}
y_{I,\nu} \equiv (Y_\nu^{\rm mass})_{II}
= \frac{\sqrt{2\,m_{\nu,I} M_I}}{v_h}.
\end{equation}
In this case the type-I seesaw relation holds pairwise,
\begin{equation}
m_{\nu,I} = \frac{v_h^2}{2}\,\frac{|y_{I,\nu}|^2}{M_I}.
\end{equation}
Outside this special aligned limit ($O\neq\mathbb{1}$ and/or a different
labelling of the heavy eigenstates), $Y_\nu^{\rm mass}$ is non-diagonal and
each $m_i$ is generated by several entries $(Y_\nu^{\rm mass})_{iI}$. One must
then retain the full matrix structure of $Y_\nu^{\rm mass}$ (equivalently,
$m_D$) in the seesaw relation, and the naive identification
$m_i = y_{i,\nu} v_h/\sqrt{2}$ for the light states is not valid.

\section{Thermal averaging and effective rates}
\label{sec:rates-vs-H}

In a radiation–dominated expanding Universe, departures of the number density of a species \(X\) from equilibrium relax back with an \emph{effective} rate that sums all number–changing channels that create or destroy \(X\), normalized to the equilibrium abundance. Linearizing the Boltzmann equation around \(n_X^{\rm eq}\),
\begin{equation}
\dot n_X + 3H n_X \;=\; -\big(n_X - n_X^{\rm eq}\big)\,\Gamma_X^{\rm eff}(T)\,,
\qquad
\Gamma_X^{\rm eff}(T) \;=\; \frac{\sum_i \gamma_i(T)}{n_X^{\rm eq}(T)}\,,
\label{eq:Gammaeff-def}
\end{equation}
where the sum runs over all relevant processes \(i\) and \(\gamma_i\) are the corresponding reaction densities (rates per unit volume). Chemical equilibrium is maintained when \(\Gamma_X^{\rm eff}\gg H\). A practical tracking criterion that accounts for the explicit temperature dependence of \(n_X^{\rm eq}\) is
\begin{equation}
\frac{\Gamma_X^{\rm eff}(T)}{H(T)} \;\gg\; \frac{3}{2} + \xi_X\,,
\qquad
\xi_X \equiv \frac{m_X}{T}\,.
\label{eq:tracking-criterion}
\end{equation}
This follows from linearizing the Boltzmann equation and demanding that the relaxation time is much shorter than the Hubble time, including the dilution term. However, in the paper we have just used the simplified condition $\frac{\Gamma_X^{\rm eff}(T)}{H(T)} \gg 1$.

\paragraph{Reaction densities and reduced cross sections.}
For a generic \(2\to 2\) process \(a b\to c d\), the reaction density is
\begin{equation}
\gamma_{ab\to cd}(T)
=
\int d\Pi_a d\Pi_b d\Pi_c d\Pi_d\,
(2\pi)^4\delta^{(4)}(p_a+p_b-p_c-p_d)\,
f_a^{\rm eq} f_b^{\rm eq}
\big(1\pm f_c^{\rm eq}\big)\big(1\pm f_d^{\rm eq}\big)\,
|\mathcal{M}_{ab\to cd}|^2,
\label{eq:gamma-phase-space}
\end{equation}
with \(d\Pi_i \equiv d^3p_i/[(2\pi)^3 2E_i]\) and
\(f^{\rm eq}_{F,B}(E)=1/(\mathrm{e}^{E/T}\pm 1)\) for fermions/bosons. Using standard manipulations and introducing the reduced cross section \(\widehat{\sigma}\equiv 2s\,\sigma\) (see, e.g., \cite{Giudice:2003jh,Davidson:2008bu,Fong:2012buy,HahnWoernle:2009qn}), this can be written for (effectively) massless initial states as
\begin{equation}
\gamma_{ab\to cd}(T)
=
\frac{T}{64\pi^4}
\int_{s_{\rm min}}^\infty ds\;
\widehat{\sigma}_{ab\to cd}(s)\,
\sqrt{s}\,
K_1\!\left(\frac{\sqrt{s}}{T}\right),
\label{eq:gamma-general}
\end{equation}
where \(K_1\) is a modified Bessel function and \(s_{\rm min}\) includes thermal masses when relevant. Quantum–statistical factors and thermal corrections are encoded in \(\widehat{\sigma}\) and in the effective threshold \(s_{\rm min}\).

When Maxwell–Boltzmann factorization holds one may also estimate \(\gamma_{ab\to cd}\simeq n_a^{\rm eq} n_b^{\rm eq}\langle\sigma v\rangle\); Eqs.~\eqref{eq:gamma-phase-space}–\eqref{eq:gamma-general} provide the exact thermal result used in our calculations.

\subsubsection*{Useful templates}

\paragraph{(i) Pair annihilation \(X X \leftrightarrow \mathrm{SM}\).}
Near equilibrium,
\begin{equation}
\dot n_X + 3H n_X = -\langle\sigma v\rangle\big(n_X^2 - n_{X}^{\rm eq\,2}\big)
\;\Rightarrow\;
\Gamma_X^{\rm eff} = 2\,\langle\sigma v\rangle\,n_X^{\rm eq}.
\label{eq:eff-ann}
\end{equation}
(The factor \(2\) counts the two \(X\) destroyed per reaction.)

\paragraph{(ii) Semi–annihilation \(X X \leftrightarrow X Y\).}
\begin{equation}
\Gamma_X^{\rm eff} = \langle\sigma v\rangle\,n_X^{\rm eq}.
\label{eq:eff-semi}
\end{equation}

\paragraph{(iii) Coannihilation \(X a \leftrightarrow \mathrm{SM}\) (partner \(a\) thermal).}
\begin{equation}
\Gamma_X^{\rm eff} \simeq \langle\sigma v\rangle_{X a}\,n_a^{\rm eq}.
\label{eq:eff-coann}
\end{equation}

\paragraph{(iv) Decay / inverse decay \(X \leftrightarrow \mathrm{SM}\).}
\begin{equation}
\Gamma_X^{\rm eff} = \Gamma_D\,\frac{K_1(\xi_X)}{K_2(\xi_X)},
\qquad
\xi_X\equiv \frac{m_X}{T},
\label{eq:eff-decay}
\end{equation}
with \(K_1/K_2\simeq \xi_X/2\) for \(\xi_X\ll 1\) and \(K_1/K_2\to 1\) for \(\xi_X\gg 1\).

\paragraph{(v) \(2\to 2\) production of a single \(X\): \(a b \to X\cdots\).}
\begin{equation}
\Gamma_X^{\rm eff} = \frac{\gamma_{ab\to X\cdots}(T)}{n_X^{\rm eq}(T)},
\label{eq:eff-2to2-singleX}
\end{equation}
with \(\gamma\) from Eq.~\eqref{eq:gamma-general}. This is the precise counterpart of the heuristic \(\Gamma\sim \sum n^{\rm eq}\langle\sigma v\rangle\).

\paragraph{(vi) Number–changing \(3\to 2\).}
\begin{equation}
\dot n_X + 3H n_X = -\langle\sigma v^2\rangle\big(n_X^3 - n_X^{\rm eq\,3}\big)
\;\Rightarrow\;
\Gamma_X^{\rm eff} = 3\,\langle\sigma v^2\rangle\,(n_X^{\rm eq})^2.
\label{eq:eff-3to2}
\end{equation}

\paragraph{(vii) Elastic (kinetic) scattering with bath species \(b\).}
\begin{equation}
\Gamma_{\rm kin} \simeq n_b^{\rm eq}\,\langle\sigma_T v\rangle,
\label{eq:eff-kin}
\end{equation}
relevant for temperature (rather than number) equilibration.

In natural units, \(n^{\rm eq}\!\sim\!\mathrm{GeV}^3\) and \(\gamma\!\sim\!\mathrm{GeV}^4\), hence \(\Gamma^{\rm eff}=\gamma/n^{\rm eq}\!\sim\!\mathrm{GeV}\). Boltzmann and threshold effects are automatically encoded in Eqs.~\eqref{eq:gamma-phase-space}–\eqref{eq:gamma-general} and in \(K_1(\xi_X)/K_2(\xi_X)\); as a result, \(\Gamma^{\rm eff}/H\) drops rapidly once \(\xi_X\gtrsim\mathcal{O}(1\text{--}3)\).

\newpage

\section{Details of $\Delta L=1$ Scatterings at High Temperature}
\label{app:DL1-scatt}

\begin{figure}
\centering

% ===================== (i) Decays and inverse decays =====================
\begin{tikzpicture}[scale=1.0]
\begin{feynman}
  % Decay: N_I -> L H
  \vertex (Ni) at (-1.8,0.0) {$N_I$};
  \vertex (v1) at (0.0,0.0) {};
  \vertex (L)  at (2.0,0.8) {$L_\alpha$};
  \vertex (H)  at (2.0,-0.8) {$H$};

  % Inverse decay: Lbar H^\dagger -> N_I
  \vertex (Lb) at (-2.0,-2.0) {$\bar L_\alpha$};
  \vertex (Hb) at ( 0.0,-3.2) {$H^\dagger$};
  \vertex (v2) at (0.0,-2.0) {};
  \vertex (Nf) at (2.0,-2.0) {$N_I$};

  \diagram{
    % N_I -> L H
    (Ni) -- [fermion] (v1),
    (v1) -- [fermion] (L),
    (v1) -- [scalar]  (H),

    % \bar L H^\dagger -> N_I
    (Lb) -- [anti fermion] (v2),
    (Hb) -- [scalar]       (v2),
    (v2) -- [fermion]      (Nf),
  };
\end{feynman}
\node at (0.1,-3.6) {(\textit{i}) Decays and inverse decays $N_I \leftrightarrow L_\alpha H,~\bar L_\alpha H^\dagger$};
\end{tikzpicture}

\vspace{1.4cm}

% ===================== (ii) ΔL=1 scatterings =====================
\begin{tikzpicture}[scale=1.0]
\begin{feynman}
  % Left: N_I L -> Q_3 t (t-channel H)
  \vertex (N1) at (-2.0, 1.2) {$N_I$};
  \vertex (L1) at (-2.0,-1.2) {$L_\alpha$};
  \vertex (a1) at (0.0, 0.6) {};
  \vertex (b1) at (0.0,-0.6) {};
  \vertex (Q3) at (2.0, 1.2) {$Q_3$};
  \vertex (t)  at (2.0,-1.2) {$t$};

  \diagram*{
    (N1) -- [fermion] (a1),
    (L1) -- [fermion] (b1),
    (a1) -- [scalar, edge label=$H$] (b1),
    (a1) -- [fermion] (Q3),
    (b1) -- [fermion] (t),
  };

  % Right: L A -> N_I H (t-channel L)
  \vertex (L2) at (4.5, 1.2) {$L_\alpha$};
  \vertex (A2) at (4.5,-1.2) {$A$};
  \vertex (a2) at (6.5, 0.6) {};
  \vertex (b2) at (6.5,-0.6) {};
  \vertex (N2) at (8.5, 1.2) {$N_I$};
  \vertex (H2) at (8.5,-1.2) {$H$};

  \diagram*{
    (L2) -- [fermion] (a2),
    (A2) -- [boson]  (b2),
    (a2) -- [fermion] (b2),
    (a2) -- [fermion] (N2),
    (b2) -- [scalar]  (H2),
  };
\end{feynman}
\node at (3.25,-2.3) {(\textit{ii}) $\Delta L=1$ scatterings: $N_I L_\alpha \leftrightarrow Q_3 t$ and $L_\alpha A \leftrightarrow N_I H$ (t-channel)};
\end{tikzpicture}

\vspace{1.6cm}

% ===================== (iii) ΔL=2 scatterings via s-channel N_I =====================
\begin{tikzpicture}[scale=1.0]
\begin{feynman}
  \vertex (L3)  at (-2.0, 1.0) {$L_\alpha$};
  \vertex (H3)  at (-2.0,-0.2) {$H$};
  \vertex (v3)  at (-0.4,0.4) {};
  \vertex (v4)  at ( 1.2,0.4) {};
  \vertex (Lb3) at ( 3.0, 1.0) {$\bar L_\beta$};
  \vertex (Hb3) at ( 3.0,-0.2) {$H^\dagger$};

  \diagram*{
    (L3)  -- [fermion] (v3),
    (H3)  -- [scalar]  (v3),
    (v3)  -- [fermion, edge label=$N_I$] (v4),
    (v4)  -- [anti fermion] (Lb3),
    (v4)  -- [scalar]       (Hb3),
  };
\end{feynman}
\node at (0.5,-1.8) {(\textit{iii}) $\Delta L=2$ scatterings: $L H \leftrightarrow \bar L H^\dagger$, $L L \leftrightarrow H H$ via s-channel $N_I$};
\end{tikzpicture}

\vspace{1.8cm}

% ===================== (iv) Portal processes with R =====================
\begin{tikzpicture}[scale=1.0]
\begin{feynman}
  % N_I N_J -> R
  \vertex (N4a) at (-2.0, 1.0) {$N_I$};
  \vertex (N4b) at (-2.0,-0.2) {$N_J$};
  \vertex (v5)  at (-0.4,0.4) {};
  \vertex (R4)  at ( 1.2,0.4) {$R$};

  \diagram*{
    (N4a) -- [fermion] (v5),
    (N4b) -- [fermion] (v5),
    (v5)  -- [scalar]  (R4),
  };

  % N_I R -> L H
  \vertex (N5) at (3.6, 1.0) {$N_I$};
  \vertex (R5) at (3.6,-0.2) {$R$};
  \vertex (v6) at (5.2,0.4) {};
  \vertex (L5) at (6.8, 1.0) {$L_\alpha$};
  \vertex (H5) at (6.8,-0.2) {$H$};

  \diagram*{
    (N5) -- [fermion] (v6),
    (R5) -- [scalar]  (v6),
    (v6) -- [fermion] (L5),
    (v6) -- [scalar]  (H5),
  };

  % R R -> H H
  \vertex (R6a) at (9.2, 1.0) {$R$};
  \vertex (R6b) at (9.2,-0.2) {$R$};
  \vertex (v7)  at (10.8,0.4) {};
  \vertex (H6a) at (12.4, 1.0) {$H$};
  \vertex (H6b) at (12.4,-0.2) {$H$};

  \diagram*{
    (R6a) -- [scalar] (v7),
    (R6b) -- [scalar] (v7),
    (v7)  -- [scalar] (H6a),
    (v7)  -- [scalar] (H6b),
  };
\end{feynman}
\node at (5.2,-1.8) {(\textit{iv}) Portal processes with $R$ (and $\chi$ via $R$): $N_I N_J \leftrightarrow R$, $N_I R \leftrightarrow L_\alpha H$, $R R \leftrightarrow H H$};
\end{tikzpicture}

\caption{
Representative neutrino-portal processes maintaining thermal contact between
the SM, the heavy neutrinos $N_I$, and the dark sector at $T\gg M_I$.}
\label{fig:neutrino-portal-processes}
\end{figure}

As discussed in the main text, for temperatures larger than the lightest heavy-neutrino mass, $T \gtrsim M_{N}$, the SM bath is in thermal contact with the hidden sector through the neutrino portal.
In this section, $H$ denotes the SM Higgs doublet in the unbroken phase, not the CP-even mass eigenstate $H$ of App.~\ref{sec:model}.

The relevant interaction terms are
\begin{align}
-\mathcal{L}_Y
&\supset
Y_\nu^{\alpha I}\,\overline{L_\alpha}\,\tilde H\,N_I
+ \frac{1}{2}\,Y_N^{IJ}\,R\,\overline{N_I^{\,c}}N_J
+ y_p\,\overline{\chi}\chi\,R
+ y_t\,\overline{Q_3}\,\tilde H\,t_R
+ \text{h.c.},
\label{eq:LY-app}\\[4pt]
\mathcal{L}_{\rm gauge}
&\supset
\sum_f \bar f\,\gamma^\mu
\big(g_2\,T^a W^a_\mu + g_Y\,Y\,B_\mu\big)f
+ (D_\mu H)^\dagger (D^\mu H)
+ \bar\chi\,i\gamma^\mu(\partial_\mu + i g_X q_\chi Z'_\mu)\chi,
\label{eq:Lgauge-app}
\end{align}
with $\tilde H = i\sigma_2 H^\ast$, $f$ running over SM fermions,
$T^a$ the $SU(2)_L$ generators, $Y$ hypercharge, and $D_\mu$ the SM
covariant derivative acting on $H$.
The process Feynman diagrams are reported in Fig.~\ref{fig:neutrino-portal-processes}.
We list below the main processes keeping the hidden sector in thermal contact with the SM:
\begin{itemize}
\item[(i)] \emph{Decays and inverse decays} (with violation of lepton number $\Delta L=1$):
\begin{equation}
N_I \;\leftrightarrow\; L_\alpha H,
\qquad
N_I \;\leftrightarrow\; \bar L_\alpha H^\dagger.
\end{equation}
These control the production and destruction of heavy neutrinos $N_I$ and are the leading
neutrino–portal processes in the relativistic and mildly non–relativistic
regimes.

\item[(ii)] \emph{$\Delta L=1$ Yukawa–mediated scatterings} with one $Y_\nu$
and one SM coupling, e.g.
\begin{equation}
N_I L_\alpha \;\leftrightarrow\; Q_3 t,
\qquad
L_\alpha A \;\leftrightarrow\; N_I H,
\end{equation}
and their crossed channels, mediated by Higgs or lepton exchange, where $A$ is a SM electroweak gauge boson. These are
$\mathcal{O}(Y_\nu^2 y_t^2)$ and $\mathcal{O}(Y_\nu^2 g^2)$ processes and
efficiently contribute to keeping $N_I$ in kinetic and chemical contact with
the SM plasma for $T\gtrsim M_I$.

\item[(iii)] \emph{$\Delta L=2$ scatterings} induced by virtual $N_I$, such as
\begin{equation}
L_\alpha H \;\leftrightarrow\; \bar L_\beta H^\dagger,
\qquad
L_\alpha L_\beta \;\leftrightarrow\; H H,
\end{equation}
which encode lepton–number violation and are relevant for washout and for the
low-energy Weinberg operator once $N_I$ are integrated out. These processes are
relevant only at very high temperatures ($T\gtrsim 10^{12}\,\mathrm{GeV}$) and
are negligible near (and below) $T\sim M_I$. We will not consider them further.

\item[(iv)] \emph{Portal processes involving the dark scalar and DM}, mediated
by $N_I$ and $R$, e.g.
\begin{equation}
N_I N_J \;\leftrightarrow\; R,
\qquad
N_I R \;\leftrightarrow\; L_\alpha H,
\qquad
R R \;\leftrightarrow\; H H,
\end{equation}
and, via the $R$–$\chi$ coupling, reactions that transfer energy and number
between $\chi$ and the SM through the $N_I$–$R$ chain. Once $N_I$ are in
equilibrium with the SM, these interactions help ensure that the dark sector is also
thermally linked.
\end{itemize}

In the following we quantify the rates of decays/inverse decays, the leading
$\Delta L=1$ scatterings, and the portal processes involving the dark scalar
and DM. We work in the unbroken phase and for $T\gg m_{\rm EW}$, taking all external
masses (except $M_{N,I}$) negligible.
When several heavy Majorana eigenstates $N_I$ are present, the slowest (last) to lose equilibrium is the \emph{lightest} one. It is therefore sufficient (and conservative)
to track a single state $N$ with mass $M_N \equiv \min_I(M_{N,I})$. In the aligned limit, we denote by $m_\nu$ the light eigenvalue paired with this lightest $N$.

\subsection{Tree-level width \texorpdfstring{$N_I\to L_\alpha H$}{N→LH}}

We start from the neutrino Yukawa interaction for a given mass eigenstate $N_I$, 
\[
\mathcal{L}_{\rm int} \supset
-\,y_{\alpha I}\,\overline{L_\alpha}\,\tilde H\,N_I +\text{h.c.}
\]
Expanding in components, this contains decays of the form $N_I \to \ell_\alpha H$. 
Summing over lepton flavours yields the total width for decays and inverse decays
\begin{align}
\Gamma_{D,I}
\equiv
\Gamma\big(N_I\to L H\big)
+ \Gamma\big(N_I\to \bar L H^\dagger\big)
=
\frac{M_{N,I}}{8\pi}\;\sum_\alpha |y_{\alpha I}|^2\,
\mathcal{F}(r_H,r_{\ell_\alpha}),
\label{eq:Gamma-total-general}
\end{align}
with
\begin{equation}
\mathcal{F}(r_H,r_{\ell_\alpha})
=
\sqrt{(1-r_H-r_{\ell_\alpha})^2-4\,r_H r_{\ell_\alpha}}\;
\big(1+r_{\ell_\alpha}-r_H\big),
\quad
r_H \equiv \frac{m_H^2}{M_{N,I}^2},
\quad
r_{\ell_\alpha} \equiv \frac{m_{\ell_\alpha}^2}{M_{N,I}^2}.
\end{equation}
In the unbroken phase, or whenever $M_{N,I}\gg m_H,m_{\ell_\alpha}$, $\mathcal{F}(r_H,r_{\ell_\alpha})\to1,$
so that
\begin{align}
\Gamma_{D,I}
&= \frac{M_{N,I}}{8\pi}\,(y^\dagger y)_{II}.
\label{eq:GammaDI-symm}
\end{align}
This is the standard expression used in leptogenesis analyses \cite{Fukugita:1986hr,Buchmuller:2004nz,Davidson:2008bu}.

\medskip

In the early Universe the heavy neutrinos $N_I$ are immersed in a thermal bath
and are produced with a momentum distribution rather than at rest. The relevant
quantity for equilibration is the \emph{time–dilated}, thermally averaged decay
rate per particle (see App.~\ref{sec:rates-vs-H}) for the lightest of the heavy
neutrinos $N$,
\begin{equation}
\langle \Gamma_{D} \rangle_T
\;=\;
\Gamma_{D}\,
\frac{K_1(\xi)}{K_2(\xi)},
\qquad
\xi \equiv \frac{M_N}{T},
\label{eq:Gamma-thermal}
\end{equation}
where $K_{1,2}$ are modified Bessel functions of the second kind and we assume
Maxwell–Boltzmann statistics for $N_I$.
To assess thermal contact we compare $\langle \Gamma_{D} \rangle_T$ with the
Hubble rate during radiation domination,
\begin{equation}
H(T)
=
1.66\,\sqrt{g_\ast(T)}\,\frac{T^2}{M_{\rm Pl}}
=
1.66\,\sqrt{g_\ast}\,
\frac{M_N^2}{M_{\rm Pl}}\,
\frac{1}{\xi^{2}},
\label{eq:Hubble-rad}
\end{equation}
where $g_\ast$ is the effective number of relativistic degrees of freedom.
Using Eqs.~\eqref{eq:Gamma-thermal} and \eqref{eq:GammaDI-symm} and in the aligned Casas–Ibarra limit ($(y^\dagger y)_{II} = 2 m_{\nu,I} M_I/v_h^2$) the ratio $\langle \Gamma_{D,I} \rangle_T/H(T)$ simplifies to
\begin{equation}
\frac{\langle \Gamma_{D} \rangle_T}{H(T)}
=
\frac{m_{\nu} M_{\rm Pl}}{4\pi\,1.66\,\sqrt{g_\ast}\,v_h^2}\;
\xi^{2}\,
\frac{K_1(\xi)}{K_2(\xi)}.
\label{eq:Gamma-over-H-mnu}
\end{equation}

We define the equilibration (or decoupling) temperature $T_\ast$ by
\begin{equation}
\left.\frac{\langle \Gamma_{D} \rangle_T}{H(T)}\right|_{T=T_\ast}
= 1.
\label{eq:Tstar-def}
\end{equation}
For $T_\ast \gg M_N$ (i.e.\ $\xi_\ast \ll 1$), we can use
$K_1(\xi)/K_2(\xi) \simeq \xi/2$, so Eq.~\eqref{eq:Gamma-over-H-mnu} yields
\begin{equation}
\frac{T_\ast}{M_N}
\simeq 2.8\,
\left(\frac{m_{\nu}}{0.05~\mathrm{eV}}\right)^{\!1/3}
\left(\frac{106.75}{g_\ast}\right)^{\!1/6},
\label{eq:Tstar-numeric}
\end{equation}
so that $N$ \emph{enter} equilibrium as the Universe cools to
$T\sim \text{few}\times M_N$, and are well in equilibrium for
$T \lesssim \text{few}\times M_N$.
Evaluating Eq.~\eqref{eq:Gamma-over-H-mnu} at $T=M_N$ ($\xi=1$) gives
\begin{equation}
\left.\frac{\langle \Gamma_{D} \rangle_T}{H(T)}\right|_{T=M_N}
=
\frac{m_{\nu} M_{\rm Pl}}{4\pi\,1.66\,\sqrt{g_\ast}\,v_h^2}\;
\frac{K_1(1)}{K_2(1)}
\sim \mathcal{O}(20\text{--}30)
\quad
\text{for } m_{\nu}\sim 0.05~\mathrm{eV},
\end{equation}
confirming that decays and inverse decays alone are fast enough to maintain
thermal contact around and below $T\sim \text{few}\times M_N$ for
seesaw–motivated parameters.

\subsection*{Top-assisted scattering $N_I L_\alpha \to Q_3 t$}

We now consider the process $N_I(p_1) + L_\alpha(p_2) \;\to\; Q_3(p_3) + t(p_4)$ mediated by $t$–channel Higgs exchange of the interaction term
\(
-\mathcal{L} \supset
y_{\alpha I}\,\overline{L_\alpha}\,\tilde H N_I
+ y_t\,\overline{Q_3} H t_R + \text{h.c.}
\)
In the relativistic regime $s\gg M_I^2,m_H^2(T)$, after spin/color sums and
angular integration one finds a reduced cross section of the form
\begin{equation}
\widehat{\sigma}_{N_I L\to Q_3 t}(s)
\;\equiv\;
2s\,\sigma_{N_I L\to Q_3 t}(s)
=
\frac{3}{8\pi}\,
(Y_\nu^\dagger Y_\nu)_{II}\,|y_t|^2\,
\mathcal{F}_t\!\left(\frac{s}{m_H^2(T)}\right),
\label{eq:sigma-hat-NLQt-def}
\end{equation}
with
\begin{equation}
\mathcal{F}_t(x)
\equiv
\left(1+\frac{2}{x}\right)\ln(1+x) - 2,
\qquad
x \equiv \frac{s}{m_H^2(T)}.
\label{eq:Ft-def}
\end{equation}
With Hard Thermal Loop (HTL) thermal masses\footnote{HTL masses are temperature-dependent effective masses that particles acquire from interactions in a hot plasma, computed within the Hard Thermal Loop approximation \cite{Braaten:1990it}.}
$m_H^2(T)\sim c_H g^2 T^2$ and typical $s\sim (2T)^2$ one has
$x=\mathcal{O}(5\text{--}20)$, so numerically
\(
\mathcal{F}_t(x)\sim 0.3\text{--}1.0
\).
A fully resummed derivation (including HTL propagators and exact
$\mathcal{F}_t$) can be found in
\cite{Giudice:2003jh,Fong:2012buy,HahnWoernle:2009qn}.

\subsection*{Gauge-assisted scattering $L_\alpha A \to N_I H$}

Now we focus on the process $L_\alpha(p_1) + A(p_2) \;\to\; N_I(p_3) + H(p_4)$ where $A$ is an electroweak gauge boson ($W_a$ or $B$).
These channels involve one neutrino Yukawa and one gauge vertex and are
representative of $\mathcal{O}(Y_\nu^2 g^2)$ scatterings.
In the unbroken phase the relevant interactions are
\begin{align}
-\mathcal{L}_Y
&\supset
Y_\nu^{\alpha I}\,\overline{L_\alpha}\,\tilde H\,N_I
+ \text{h.c.},\\
\mathcal{L}_{\rm gauge}
&\supset
g\,\bar L_\alpha \gamma^\mu T^a L_\alpha\,A^a_\mu
+ (D_\mu H)^\dagger(D^\mu H),
\end{align}
with analogous hypercharge couplings.
A representative contribution is the $t$–channel lepton exchange; $s$– and
$u$–channel diagrams complete the gauge–invariant set and yield the same
parametric dependence once summed.

Summing over spins, polarizations and gauge indices, and working in the
relativistic regime $s\gg M_I^2,m_{L,H,A}^2(T)$, one obtains the reduced
cross section
\begin{equation}
\widehat{\sigma}_{L A\to N_I H}(s)
\equiv
2s\,\sigma_{L A\to N_I H}(s)
=
\frac{C_A}{8\pi}\,
|Y_\nu^{\alpha I}|^2\,g^2\,
\mathcal{F}_g\!\left(\frac{s}{m_{\rm th}^2(T)}\right),
\label{eq:sigma-hat-LA-NH}
\end{equation}
where: $C_A=\mathcal{O}(1)$ is a group-theory factor, $m_{\rm th}^2(T)$ is a representative thermal scale built from $m_L^2(T),m_H^2(T),m_A^2(T)$, and $\mathcal{F}_g$ is a dimensionless function encoding angular structure and IR regularization.
A convenient parametrization with the correct limits is
\begin{equation}
\mathcal{F}_g(x)
\equiv
\left(1+\frac{1}{x}\right)\ln(1+x) - 1,
\qquad
x \equiv \frac{s}{m_{\rm th}^2(T)},
\label{eq:Fg-def}
\end{equation}
For $m_{\rm th}^2(T)\sim c\,g^2 T^2$ and $s\sim(2T)^2$ one typically has
$x=\mathcal{O}(5\text{--}30)$ and hence
$\mathcal{F}_g(x)\sim 0.5\text{--}2$.
Summing over flavours,
\begin{equation}
\widehat{\sigma}_{L A\to N_I H}(s)
=
\frac{C_A}{8\pi}\,
(Y_\nu^\dagger Y_\nu)_{II}\,g^2\,
\mathcal{F}_g\!\left(\frac{s}{m_{\rm th}^2(T)}\right).
\label{eq:sigma-hat-LA-NH-final}
\end{equation}
Detailed HTL–resummed expressions and precise values of $\mathcal{F}_g$ for the
individual gauge channels can be found in
\cite{Giudice:2003jh,Fong:2012buy,HahnWoernle:2009qn}.

\subsection*{Per-particle rates and comparison with Hubble rate for the top- and Gauge-assisted scatterings}

For equilibration we require the \emph{per–particle} rate, normalized to one
heavy neutrino $N_I$:
\begin{equation}
\Gamma_{\rm scatt}^{(\Delta L=1)}(T)
\equiv
\frac{\gamma_{\Delta L=1}(T)}{n_{N_I}^{\rm eq}(T)},
\end{equation}
where $\gamma_{\Delta L=1}$ is the sum of the relevant $\Delta L=1$ reaction
densities involving $N_I$. In the relativistic limit ($T\gg M_I$), the
equilibrium number density of a Majorana fermion ($g_N=2$) is
\begin{equation}
n_{N_I}^{\rm eq}(T)
=
\frac{3\,\zeta(3)}{2\pi^2}\,T^3.
\end{equation}

For a slowly varying reduced cross section
$\widehat{\sigma}(s)\approx\widehat{\sigma}_0$,
Eq.~\eqref{eq:gamma-general} gives
\begin{equation}
\gamma(T)
\simeq
\frac{\widehat{\sigma}_0\,T^4}{16\pi^4},
\end{equation}
using $\int_0^\infty dx\,x^2 K_1(x)=2$.
Thus
\begin{equation}
\Gamma_{\rm scatt}(T)
=
\frac{\gamma(T)}{n_{N_I}^{\rm eq}(T)}
\simeq
\frac{\widehat{\sigma}_0}{24\,\zeta(3)\,\pi^2}\,T.
\label{eq:Gamma-from-sigmahat}
\end{equation}

The scattering rate for top-assisted scatterings can be determined using Eq.~\eqref{eq:sigma-hat-NLQt-def}:
\begin{equation}
\Gamma^{(\Delta L=1)}_{{\rm top},I}(T)
\simeq
\frac{3\,\overline{\mathcal{F}}_t}{192\,\zeta(3)\,\pi^3}\,
(Y_\nu^\dagger Y_\nu)_{II}\,y_t^2\,T
\equiv
c_t\,(Y_\nu^\dagger Y_\nu)_{II}\,y_t^2\,T,
\end{equation}
with $c_t = 3\,\overline{\mathcal{F}}_t/(192\,\zeta(3)\,\pi^3) \approx (2\text{--}4)\times 10^{-4}$,
which parametrizes the theoretical uncertainties associated with thermal masses,
IR regularization and phase–space approximations.
Instead, for gauge-assisted processes and using Eq.~\eqref{eq:sigma-hat-LA-NH-final},
we find
\begin{equation}
\Gamma^{(\Delta L=1)}_{{\rm gauge},I}(T)
\simeq
\frac{C_A\,\overline{\mathcal{F}}_g}{192\,\zeta(3)\,\pi^3}\,
(Y_\nu^\dagger Y_\nu)_{II}\,g^2\,T
\equiv
c_g\,(Y_\nu^\dagger Y_\nu)_{II}\,g^2\,T,
\end{equation}
with $c_g = C_A\,\overline{\mathcal{F}}_g/(192\,\zeta(3)\,\pi^3) \approx
(1\text{--}3)\times10^{-4}$.

Summing the dominant channels,
\begin{equation}
\Gamma_{\rm scatt}^{(\Delta L=1)}(T)
\equiv
\Gamma^{(\Delta L=1)}_{{\rm top},I}
+\Gamma^{(\Delta L=1)}_{{\rm gauge},I}
\simeq
c_1\,
(Y_\nu^\dagger Y_\nu)_{II}\,(y_t^2+g^2)\,T,
\label{eq:Gamma-scatt-scaling-refined}
\end{equation}
with an effective coefficient
\begin{equation}
c_1 \approx (3\text{--}7)\times 10^{-4},
\end{equation}
once thermal masses, angular dependence and channel multiplicities are taken
into account, in agreement with detailed numerical studies
\cite{Giudice:2003jh,HahnWoernle:2009qn,Fong:2012buy}.
The variation in the parameters $c_t$, $c_g$ and consequently $c_1$
parametrizes the theoretical uncertainty from thermal masses, IR regularization
and phase–space approximations in the $\Delta L=1$ scatterings.

In the aligned limit $\Gamma_{\rm scatt}^{(\Delta L=1)}(T)$ can be rewritten as
\begin{equation}
\Gamma_{\rm scatt}^{(\Delta L=1)}(T)
\simeq
2c_1\,
\frac{m_{\nu,I} M_I}{v_h^2}\,
(y_t^2+g^2)\,T.
\end{equation}
Therefore, the ratio between the scattering and the Hubble rate is 
\begin{equation}
\frac{\Gamma_{\rm scatt}^{(\Delta L=1)}(T)}{H(T)}
\simeq
\frac{2c_1}{1.66\sqrt{g_\ast}}\,
\frac{m_{\nu,I} M_{\rm Pl}}{v_h^2}\,
(y_t^2+g^2)\,
\frac{M_I}{T}.
\end{equation}
For $m_{\nu,I}\sim 0.05~\mathrm{eV}$, $g_\ast\sim 10^2$,
$y_t^2+g^2=\mathcal{O}(1)$ and $c_1\sim 5 \times  10^{-4}$ one obtains for $T\sim M_I$
\begin{equation}
\frac{\Gamma_{\rm scatt}^{(\Delta L=1)}}{H}
\gtrsim \mathcal{O}(1).
\end{equation}

\subsection{Portal processes involving $R$ and $\chi$}

\textbf{(a) $R R \leftrightarrow H H$.} The cross section of the process $R R \leftrightarrow H H$ depends on $\kappa_{\rm loop}$. In fact, for $T\gg m_{H},m_{R}$ it scales as
\begin{equation}
\sigma(R R \to H H) \sim \frac{\kappa_{\rm loop}^2}{16\pi\,s},
\qquad s\sim T^2.
\end{equation}
The per–particle rate is
\begin{equation}
\Gamma_{R R\to H H} \sim n_R^{\rm eq}\,\sigma \sim
\kappa_{\rm loop}^2\,T,
\end{equation}
up to $\mathcal{O}(10^{-1})$ factors. For the values relevant in our model,
$\kappa_{\rm loop}\sim 10^{-11}-10^{-10}$, hence
$\Gamma_{R R\to H H}/T\sim10^{-20}$ and $\Gamma_{R R\to H H}\ll H(T)$ at all
temperatures of interest.
Thus $R R\leftrightarrow H H$ is completely negligible for thermal contact.

\medskip
\noindent
\textbf{(b) $N_I R \to L H$.}
The process $N_I R \to L_\alpha H$ proceeds through the combination of the
Majorana coupling $R N_I N_I$ and the Dirac Yukawa $L_\alpha H N_I$. In the
mass basis (and in the limit where $R$ couples diagonally to $N_I$), the
relevant interactions are
\begin{equation}
-\mathcal{L} \supset
\frac{1}{2}\,y_{N,I}\,R\,\overline{N_I^{\,c}}N_I
+ y_{\alpha I}\,\overline{L_\alpha}\,\tilde H\,N_I
+ \text{h.c.},
\end{equation}
so that $N_I R \to L_\alpha H$ proceeds dominantly via $s$–channel $N_I$
exchange. In the unbroken phase and for $T\gg m_{\rm EW}$, taking all external
masses (except $M_I$) negligible, the reduced cross section
$\widehat{\sigma}(s)\equiv 2s\,\sigma(s)$ is
\begin{equation}
\widehat{\sigma}_{N_I R\to L H}(s)
=
\frac{1}{8\pi}\,
|y_{N,I}|^2\,
(Y_\nu^\dagger Y_\nu)_{II}\,
\mathcal{F}_{NR}\!\left(\frac{s}{M_I^2}\right),
\label{eq:sigmahat-NR}
\end{equation}
where $\mathcal{F}_{NR}$ is a dimensionless kinematic function that encodes the
exact angular dependence and the mild $s/M_I^2$ dependence. For
$s\gtrsim (2T)^2 \gg M_I^2$ one has
\begin{equation}
\mathcal{F}_{NR}(x)\xrightarrow[x\gg1]{} 1
\quad\text{up to }\mathcal{O}\!\left(\frac{M_I^2}{s}\right)\text{ corrections}.
\end{equation}
We therefore take an effective thermal average
$\overline{\mathcal{F}}_{NR}\sim 0.5$--$2$ for $T\sim\text{few}\times M_I$.

Using Eq.~\eqref{eq:gamma-general} the per–particle rate is
\begin{equation}
\Gamma_{N_I R\to L H}(T)
\equiv
\frac{\gamma_{N_I R\to L H}}{n_{N_I}^{\rm eq}}
\simeq
\frac{\widehat{\sigma}_0}{24\,\zeta(3)\,\pi^2}\,T
=
c_{NR}\,
|y_{N,I}|^2\,
(Y_\nu^\dagger Y_\nu)_{II}\,
T,
\label{eq:Gamma-NR-Yuk}
\end{equation}
with
\begin{equation}
c_{NR}
=
\frac{\overline{\mathcal{F}}_{NR}}{192\,\zeta(3)\,\pi^3}
\;\approx\;
(0.7\text{--}2.8)\times10^{-4}.
\end{equation}
This is directly analogous to the $c_t$ and $c_g$ factors obtained for the
top– and gauge–assisted $\Delta L=1$ scatterings.

In the aligned Casas–Ibarra limit one finds
\begin{equation}
\Gamma_{N_I R\to L H}(T)
\simeq
2\,c_{NR}\,
\frac{y_{N,I}^2\,m_{\nu,I}\,M_I}{v_h^2}\,T.
\label{eq:Gamma-NR-yN}
\end{equation}
For illustration, take
$m_{\nu,I}=0.05~\mathrm{eV}$,
$v_h=246~\mathrm{GeV}$,
$g_\ast=106.75$,
and $c_{NR}\sim 10^{-4}$. Using Eq.~\eqref{eq:Gamma-NR-yN}, the rate at
$T\simeq M_I$ is
\begin{equation}
\frac{\Gamma_{N_I R\to L H}}{H}
\simeq
0.12\;
\left(\frac{c_{NR}}{10^{-4}}\right)
\left(\frac{y_{N,I}}{1}\right)^2
\left(\frac{m_{\nu,I}}{0.05~\mathrm{eV}}\right)
\left(\frac{106.75}{g_\ast}\right)^{1/2},
\end{equation}
independent of $M_I$ at $T=M_I$. Thus, for $y_{N,I}\sim 1$ this channel
contributes at the level $\Gamma/H\sim 0.1$ near $T\simeq M_I$, while
larger Yukawas $y_{N,I}\gtrsim 3$ would make it fully efficient
($\Gamma/H\gtrsim 1$). For fixed $M_I$, the ratio scales as $\Gamma/H
\propto M_I/T$ at higher temperatures, decreasing slowly for $T\gg M_I$,
analogously to the standard $\Delta L=1$ scatterings.

\subsubsection{Summary}

\paragraph*{Process hierarchy and temperature trends.}
For seesaw–motivated parameters ($m_{\nu,I}\sim0.05~\mathrm{eV}$,
$y_{N,I}=\mathcal{O}(1)$, $M_I\gtrsim\text{few TeV}$, $g_\ast\simeq 10^2$):

\begin{itemize}
\item \textbf{Decays/inverse decays} $N_I\leftrightarrow L H$:
\[
\frac{\langle \Gamma_{D,I} \rangle_T}{H}
=
\frac{m_{\nu,I} M_{\rm Pl}}{4\pi\,1.66\,\sqrt{g_\ast}\,v_h^2}\;
\xi^{2}\,\frac{K_1(\xi)}{K_2(\xi)}.
\]
For $\xi\ll1$ (very high $T$), this scales as $\propto \xi^3$ and is small;
as the Universe cools to $\xi\sim\mathcal{O}(1)$ ($T\sim M_I$) the ratio grows
to $\mathcal{O}(10\text{--}30)$ and \emph{dominates} equilibration.

\item \textbf{$\Delta L=1$ scatterings} (top– and gauge–assisted):
\[
\frac{\Gamma_{\rm scatt}^{(\Delta L=1)}}{H}
\simeq
\frac{2c_1}{1.66\sqrt{g_\ast}}\,
\frac{m_{\nu,I} M_{\rm Pl}}{v_h^2}\,
(y_t^2+g^2)\,
\frac{M_I}{T}
\;\propto\; \frac{M_I}{T}.
\]
They are typically $\mathcal{O}(1)$ at $T\sim M_I$ and provide efficient
additional channels; for $T\gg M_I$ the ratio decreases slowly as $M_I/T$.

\item \textbf{Portal scattering} $N_I R\leftrightarrow L_\alpha H$:
at $T=M_I$
\begin{equation}
\left.\frac{\Gamma_{N_I R\to L H}}{H}\right|_{T=M_I}
\simeq
0.1\text{--}0.2\,
\left(\frac{c_{NR}}{10^{-4}}\right)
\left(\frac{m_{\nu,I}}{0.05~\mathrm{eV}}\right)
\left(\frac{y_{N,I}}{1}\right)^2,
\end{equation}
where the numerical range reflects the mild dependence on $g_\ast$.
For $y_{N,I}=\mathcal{O}(1)$ and $m_{\nu,I}$ in the experimentally
favored range, the portal scattering is \emph{subdominant} compared to
decays and the standard $\Delta L=1$ scatterings, but still non-negligible.
For larger Yukawas $y_{N,I}\gtrsim 3$ it can become fully efficient
($\Gamma/H\gtrsim 1$), further assisting SM–dark-sector thermal contact.

\item \textbf{Loop–induced scalar portal} $R R\leftrightarrow H H$:
$\Gamma/H \ll 1$ throughout the parameter space of interest and can be
safely neglected for thermal contact.
\end{itemize}

\noindent
\textbf{Overall:} As the Universe cools down to $T\sim \text{few}\times M_I$,
$N_I$ decays/inverse decays become the \emph{dominant} equilibration channel.
$\Delta L=1$ scatterings are efficient and comparable to decays around
$T\sim M_I$, while $N_I R\leftrightarrow L H$ provides an additional but
subleading portal for $y_{N,I}\sim 1$, becoming important only for larger
$y_{N,I}$. At $T\ll M_I$ all $N_I$–mediated reactions decouple due to Boltzmann
suppression of $N_I$.

\section{Radiative generation of the mixed quartic $(\Phi^\dagger\Phi)(R^\dagger R)$}
\label{sec:radiative}

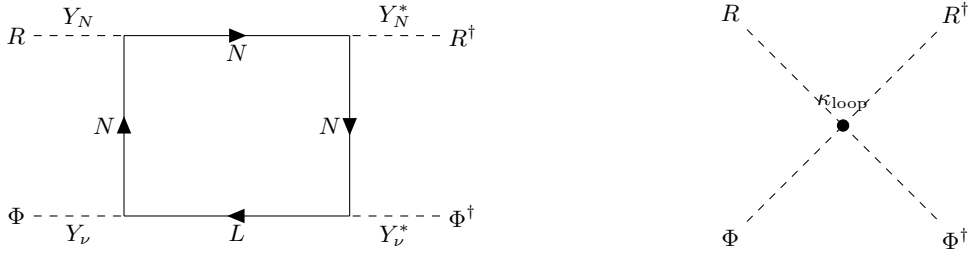
\begin{figure}[t]
\centering

% Full one-loop box
\begin{tikzpicture}[baseline={(current bounding box.center)}]
\begin{feynman}
  % corners of the loop (clockwise)
  \vertex (A) at (0,  1.2);
  \vertex (B) at (3.0, 1.2);
  \vertex (C) at (3.0,-1.2);
  \vertex (D) at (0, -1.2);

  % external scalars
  \vertex[left=1.2cm of A] (PhiL) {$R$};
  \vertex[right=1.2cm of B] (PhiR) {$R^\dagger$};
  \vertex[right=1.2cm of C] (HbotR) {$\Phi^\dagger$};
  \vertex[left=1.2cm of D] (HbotL) {$\Phi$};

  \diagram*{
    (A) -- [fermion, edge label'=$N$] (B)
        -- [fermion, edge label'=$N$] (C)
        -- [fermion, edge label=$L$]   (D)
        -- [fermion, edge label=$N$]   (A),

    (PhiL)  -- [scalar, edge label=$Y_N$] (A),
    (PhiR)  -- [scalar, edge label'=$Y_N^\ast$] (B),
    (HbotR) -- [scalar, edge label=$Y_\nu^\ast$] (C),
    (HbotL) -- [scalar, edge label'=$Y_\nu$] (D),
  };
\end{feynman}
\end{tikzpicture}
\hspace{2.8cm}
% EFT contact operator
\begin{tikzpicture}[baseline={(current bounding box.center)}]
\begin{feynman}
  \vertex (v) at (0,0) [dot,label=above:$\kappa_{\rm loop}$] {};

  \vertex[above left=1.5cm and 1.5cm of v] (PhiL) {$R$};
  \vertex[above right=1.5cm and 1.5cm of v] (PhiR) {$R^\dagger$};
  \vertex[below left=1.5cm and 1.5cm of v] (HL) {$\Phi$};
  \vertex[below right=1.5cm and 1.5cm of v] (HR) {$\Phi^\dagger$};

  \diagram*{
    (PhiL) -- [scalar] (v),
    (PhiR) -- [scalar] (v),
    (HL)   -- [scalar] (v),
    (HR)   -- [scalar] (v),
  };
\end{feynman}
\end{tikzpicture}

\caption{One-loop generation of the mixed quartic $(\Phi^\dagger \Phi)(R^\dagger R)$.
Left: box diagram with two $Y_N$ and two $Y_\nu$ insertions and internal $N,L$ lines.
Right: effective contact interaction with coupling $\kappa_{\rm loop}$ obtained by
integrating out the heavy neutrinos for $p^2 \ll M_N^2$.}
\label{fig:correct-box}
\end{figure}

In this appendix we show how the mixed quartic
$\kappa\,(\Phi^\dagger\Phi)(R^\dagger R)$ is generated at one loop from the
neutrino Yukawa interactions if it is absent (or highly suppressed) at tree
level. We focus on the part of the Lagrangian relevant for the loop, namely
the Yukawa terms,
\begin{equation}
\mathcal{L} \supset
-\,Y_\nu^{\alpha i}\,\overline{L_\alpha}\,\tilde{\Phi}\,N_i
-\frac{1}{2}\,Y_N^{ij}\,R\,\overline{N_i^{\,c}}N_j
+\text{h.c.},
\label{eq:app-L-int}
\end{equation}
and assume that the tree-level portal
\(\kappa_0(\Phi^\dagger\Phi)(R^\dagger R)\) vanishes in the UV (e.g.\ due to an
approximate symmetry or sequestering). The leading contribution to $\kappa$ is
then radiative, generated by the box diagram in the left panel of
Fig.~\ref{fig:correct-box}, which induces an effective
$(\Phi^\dagger\Phi)(R^\dagger R)$ interaction once the heavy neutrinos are
integrated out (right panel). For clarity, we first consider a single
generation: one lepton doublet $L$, one Majorana fermion $N$ with mass $M_N$,
and couplings $Y_\nu$ and $Y_N$ (the three-flavour case is given at the end).

The Feynman rule for the vertex $RNN$ is $i\,Y_N\,P_R \;-\; i\,Y_N^\ast\,P_L$,
while for $\Phi L N$ it is $-i\,Y_\nu\,P_R$. We match the four-point function
\(\Phi R \to \Phi^\dagger R^\dagger\) at vanishing external momenta. In the
regime $p_{\rm ext}^2\ll M_N^2$, the amplitude admits a local expansion and
the leading term is the Wilson coefficient of $(\Phi^\dagger\Phi)(R^\dagger R)$.
Higher orders in $p_{\rm ext}^2/M_N^2$ map onto derivative operators and are
irrelevant for $\kappa$. In our setup $M_{N_i}\gtrsim 10$~TeV, so for all
phenomenological applications $p^2/M_{N_i}^2\ll 1$ and the zero-momentum
matching is an excellent approximation.

\medskip

\paragraph{Single generation: loop evaluation.}
Choosing a fermion-flow orientation along the loop, the box gives
\begin{equation}
i\mathcal{M}
=
(-1)\,(-iY_N)(-iY_N^\ast)(-iY_\nu)(-iY_\nu^\ast)
\int\!\frac{d^d k}{(2\pi)^d}\;
\mathrm{Tr}\!\left[
P_R S_N(k)\,P_L S_N(k)\,P_L S_L(k)\,P_R S_N(k)
\right],
\label{eq:app-amp-start-full}
\end{equation}
where the overall minus sign is for the closed fermion loop. Inserting
propagators,
\begin{equation}
i\mathcal{M}
=
-\,|Y_N|^2 |Y_\nu|^2
\int\!\frac{d^d k}{(2\pi)^d}\;
\frac{\mathcal{N}(k)}{(k^2 - M_N^2)^3\,k^2},
\end{equation}
with
\begin{equation}
\mathcal{N}(k)
=
\mathrm{Tr}\!\big[
P_R(\slashed{k}+M_N)\,
P_L(\slashed{k}+M_N)\,
P_L\slashed{k}\,
P_R(\slashed{k}+M_N)
\big]
= 2M_N^2 k^2.
\end{equation}
Thus
\begin{equation}
i\mathcal{M}
=
-\,|Y_N|^2 |Y_\nu|^2\,2 M_N^2
\int\!\frac{d^d k}{(2\pi)^d}\,
\frac{1}{(k^2 - M_N^2)^3}.
\label{eq:app-amp-mid-full}
\end{equation}
In dimensional regularization ($d=4-2\epsilon$),
\begin{equation}
I_3(M_N)
\equiv
\mu^{2\epsilon}
\int\!\frac{d^d k}{(2\pi)^d}\,
\frac{1}{(k^2 - M_N^2)^3}
=
\frac{i}{16\pi^2}\,\frac{1}{2 M_N^2}
\left[
\frac{1}{\epsilon}
- \gamma_E + \ln(4\pi)
+ \ln\frac{\mu^2}{M_N^2}
+ 1
\right],
\end{equation}
whence
\begin{equation}
i\mathcal{M}
=
-\,i\,\frac{|Y_N|^2 |Y_\nu|^2}{16\pi^2}
\left[
\frac{1}{\epsilon}
- \gamma_E + \ln(4\pi)
+ \ln\frac{\mu^2}{M_N^2}
+ 1
\right].
\label{eq:app-amp-div-full}
\end{equation}

The low-energy EFT contains
\(
\mathcal{L}_{\rm EFT} \supset
-\,\kappa(\mu)\,(\Phi^\dagger\Phi)(R^\dagger R)
\Rightarrow
i\mathcal{M}_{\rm EFT}=-i\kappa_{\rm{loop}}(\mu)
\).
Matching in $\overline{\mathrm{MS}}$ and subtracting
$\tfrac{1}{\epsilon}-\gamma_E+\ln4\pi$ yields
\begin{equation}
\kappa_{\rm{loop}}(\mu)
=
\frac{|Y_N|^2 |Y_\nu|^2}{16\pi^2}
\left[
\ln\frac{M_N^2}{\mu^2} - 1
\right].
\label{eq:kappa-1gen-mu}
\end{equation}
Choosing $\mu \simeq M_N$ minimizes the logarithm and gives
\begin{equation}
\kappa_{\rm loop}(M_N)
\;=\;
-\frac{|Y_\nu|^{2}\,|Y_N|^{2}}{16\pi^2}.
\label{eq:kloop1gen}
\end{equation}
Using the one-generation seesaw relation (aligned limit),
\(
(Y_\nu^\dagger Y_\nu)=2m_\nu M_N/v_h^2
\),
Eq.~\eqref{eq:kappa-1gen-mu} can be written as
\begin{equation}
\kappa_{\rm loop}(\mu)
=
-\,\frac{y_{N}^{2}\,M_N}{8\pi^2 v_h^2}\;m_{\nu}
\;+\;
\frac{m_\nu\,y_N^2\,M_N}{8\pi^2 v_h^2}\,
\ln\!\frac{M_N^2}{\mu^2},
\qquad
\Rightarrow\quad
\kappa_{\rm loop}(M_N)
=
-\,\frac{y_{N}^{2}\,M_N}{8\pi^2 v_h^2}\;m_{\nu}.
\label{eq:kloop-mnu-matching}
\end{equation}
Here $y_N$ denotes the (real) eigenvalue of $Y_N$ in the one-generation case.
The sign of the loop-induced $\kappa$ is scheme-dependent, but physical observables such as the mixing angle $\alpha$ and the scalar masses are RG- and scheme-independent; only these combinations have physical meaning.

\medskip

\paragraph{Three generations and the choice of \texorpdfstring{$\mu$}{mu}.}
For three generations and in the Casas--Ibarra aligned limit
($O=\mathbb{1}$) one has
\(
(Y_\nu^\dagger Y_\nu)_{II}=2 m_{\nu,I} M_I / v_h^2
\).
Generalizing Eq.~\eqref{eq:kappa-1gen-mu} to three flavours and working in the
heavy-neutrino mass basis gives
\begin{equation}
\kappa_{\rm loop}(\mu)
=
-\sum_{I=1}^3
\frac{y_{N,I}^2\,M_I}{8\pi^2 v_h^2}\;m_{\nu,I}
\;+\;
\sum_{I=1}^3
\frac{y_{N,I}^2\,M_I}{8\pi^2 v_h^2}\;m_{\nu,I}\,
\ln\!\frac{M_I^2}{\mu^2},
\qquad
\Rightarrow\quad
\kappa_{\rm loop}\big|_{\mu=M_I\;\text{(stepwise)}}
=
-\sum_{I=1}^3
\frac{y_{N,I}^2\,M_I}{8\pi^2 v_h^2}\;m_{\nu,I}.
\label{eq:kloop-simple}
\end{equation}
With a \emph{single} renormalization scale \(\mu\), one can cancel the
logarithm for \emph{at most one} heavy mass; it is not possible to choose
\(\mu=M_I\) for all loops simultaneously if the $M_I$ are non-degenerate. Only
if $M_1\simeq M_2\simeq M_3\equiv M$ can a single choice \(\mu=M\) make all
logs small. The correct multi-scale treatment is stepwise EFT matching:
integrate out each $N_I$ at \(\mu\simeq M_I\) (so its log vanishes and leaves
the finite threshold $-1$), then run \(\kappa\) between thresholds.

\section{Renormalization-Group Evolution of the Portal Coupling \texorpdfstring{$\kappa$}{kappa}}
\label{app:kappa-running}

In this appendix we clarify how the Higgs--singlet portal coupling
\(\kappa\) evolves under renormalization-group (RG) running once the
tree-level value \(\kappa(\Lambda)=0\) is imposed at the UV scale
\(\Lambda\), and why the loop-induced contribution generated at the
heavy-neutrino threshold \(M_N\) remains extremely small at all lower
energies.

We assume that the scalar portal
\begin{equation}
\kappa(\Lambda)\,(\Phi^\dagger\Phi)(R^\dagger R)
\end{equation}
is absent at the UV scale \(\Lambda\), due to sequestering, a symmetry,
or a non-abelian UV origin of the dark gauge group \cite{ArkaniHamed:1998rs,Randall:1999ee,Luty:2001zv,Randall:1998uk,Dienes:1996zr,Holdom:1985ag,Feldman:2007wj,Pospelov:2008jd}. Since in our model
no light degrees of freedom carry simultaneously SM and dark charge,
no tree-level operator can regenerate \(\kappa\) at lower scales. The
leading nonzero contribution arises at the scale where the heavy
neutrinos \(N_I\) are integrated out. As shown in
Sec.~\ref{sec:radiative}, the one-loop box diagram involving the
neutrino Yukawas \(Y_\nu\) and \(Y_N\) generates the effective quartic
\begin{equation}
\Delta\kappa \equiv \kappa_{\rm loop}
\simeq -\sum_{I=1}^3
\frac{y_{N,I}^2\,M_{N,I}}{8\pi^2 v_h^2}\; m_{\nu,I},
\end{equation}
which is proportional to the light neutrino masses \(m_{\nu,I}\).
Numerically, for \(m_{\nu,I}\sim 0.05~\text{eV}\), \(y_{N,I}\sim 1\),
and \(M_{N,I}\sim {\cal O}(10)\,\text{TeV}\), one finds
\(|\kappa_{\rm loop}|\sim 10^{-13}\)–\(10^{-12}\), with only mild
variations in the multi–TeV range. This value serves as the matching
condition at \(\mu\simeq M_N\).

Below the heavy-neutrino scale, the effective theory contains only
\(\Phi\), \(R\), \(\chi\), and SM fields. Crucially, after the heavy
neutrinos are integrated out there are \emph{no} light fields
simultaneously charged under the SM and the dark sector that can source
a new \((\Phi^\dagger\Phi)(R^\dagger R)\) operator. As a result, the
beta function for \(\kappa\) has no additive term and takes the
multiplicative form
\begin{equation}
\beta_\kappa(\mu) \equiv \frac{d\kappa}{d\ln\mu}
= \kappa(\mu)\;\mathcal{F}(\lambda_H,\lambda_R,y_p,g_i,\dots),
\end{equation}
where $\mathcal{F}$ is a polynomial in the low-energy couplings. The
solution is
\begin{equation}
\kappa(\mu)
= \kappa(M_N)\,
\exp\!\left[
\int_{\mu}^{M_N} d\ln\mu'\;\mathcal{F}(\mu')
\right].
\label{eq:kappa-RGE-solution-short}
\end{equation}
Thus the RG evolution of \(\kappa\) is purely multiplicative
(logarithmic): the origin \(\kappa=0\) is an RG fixed point. In
particular, if the portal vanishes at the matching scale,
\begin{equation}
\kappa(M_N)=0
\quad\Longrightarrow\quad
\kappa(\mu)=0
\qquad\forall\;\mu<M_N,
\end{equation}
and RG running cannot generate a nonzero tree-level portal at lower
energies. Physically, this reflects the absence of particles carrying
both SM and dark quantum numbers: once the heavy neutrinos are removed
from the spectrum, no loop diagram can ``reconnect'' $\Phi$ with $R$
unless the portal is already present.

If instead the portal is generated radiatively at the threshold,
\(\kappa(M_N)=\kappa_{\rm loop}\neq 0\), then Eq.~\eqref{eq:kappa-RGE-solution-short}
shows that \(\kappa(\mu)\) simply undergoes multiplicative running,
\begin{equation}
\kappa(\mu)
= \kappa_{\rm loop}\,
\exp\!\left[ \int_{\mu}^{M_N} \! d\ln\mu'\,
   \mathcal{F}(\mu') \right].
\end{equation}
The exponent is controlled by SM quartics, Yukawas, and gauge couplings
and is at most \({\cal O}(1)\) between the electroweak scale and
\(M_N\), leading to an overall variation by at most a factor of a few,
\(\kappa(\mu)\sim (1\text{--}10)\,\kappa_{\rm loop}\). Therefore, in
the neutrino-aligned scenario considered in this work, the portal
coupling remains extremely small along the entire RG flow, with a
typical upper size
\begin{equation}
\kappa(\mu) \;\lesssim\; \mathcal{O}(10^{-11})
\qquad \text{for all } \mu < M_N,
\end{equation}
unless one adds new messenger fields with sizeable SM and dark charges
or abandons the seesaw-motivated parameter regime.

\section{Decoupling and freeze--out with two temperatures ($T'\neq T$)}
\label{app:2T-freezeout}

This appendix collects the ingredients used in the main text to describe the
decoupling of the neutrino portal and the subsequent dark--sector freeze--out in
the regime where the SM bath and the hidden sector evolve with different
temperatures $T'\neq T$. We adopt the notation of App.~\ref{app:DL1-scatt}:
$M_I$ denotes the heavy--neutrino masses, $\zeta\equiv T'/T$ is the dark--to--visible
temperature ratio, $g_{*}^{\rm vis}(T)$ and $g_{*S}^{\rm vis}(T)$ are the usual
visible--sector energy and entropy relativistic d.o.f., with analogous
definitions for the dark sector.

\subsection{Decoupling around $T\sim M_I$ and entropy bookkeeping}

As the temperature drops below the mass of the lightest heavy neutrino $M_{N,I}$,
the heavy--neutrino abundance becomes Boltzmann suppressed and portal--mediated
energy exchange shuts off. From that moment on, the comoving entropies of the
two sectors are separately conserved,
\begin{equation}
s_{\rm vis}(T)\,a^3=\text{const},
\qquad
s_{\rm dark}\!\big(T',\{\mu_i(T')\}\big)\,a^3=\text{const},
\label{eq:2T_entropy_cons}
\end{equation}
where $\mu_i$ is the chemical potential for the particle species $i$.
The temperature ratio at later times is fixed by separate entropy conservation after
decoupling. Assuming full chemical equilibrium in the dark sector at the
decoupling temperature $T'_{\rm dec}$ ($\mu_i(T'_{\rm dec})=0$), the general
expression is (see, e.g., \cite{Bringmann:2020mgx})
\begin{equation}
\zeta(T)\;\equiv\;\frac{T'}{T} \;=\;
\left[\frac{s_{\rm vis}(T)}{s_{\rm vis}(T_{\rm dec})}\right]^{\!1/3}
\left[\frac{s_{\rm dark}(T'_{\rm dec},\{\mu_i{=}0\})}{s_{\rm dark}(T'=\zeta T,\{\mu_i(T')\})}\right]^{\!1/3}.
\label{eq:xi-entropy-master}
\end{equation}

If the following two physical conditions apply:
(i) the relevant species in each sector behave as a radiation bath so that
\(s=\tfrac{2\pi^2}{45}g_{*S}T^3\), and
(ii) \emph{number--changing} reactions are fast enough to enforce \(\mu_i=0\),
Eq.~\eqref{eq:xi-entropy-master} reduces to the familiar result (see e.g.\ \cite{Bringmann:2020mgx})
\begin{equation}
\zeta(T)
=\left[\frac{g_{*S}^{\rm vis}(T)}{g_{*S}^{\rm vis}(T_{\rm dec})}\right]^{1/3}
 \left[\frac{g_{*S}^{\rm dark}(T'_{\rm dec})}{g_{*S}^{\rm dark}(T')}\right]^{1/3}.
\label{eq:xi_gstar_limit_app}
\end{equation}
In typical benchmarks with $m_\chi\sim\mathcal{O}({\rm GeV})$ and
$M_I\sim\mathcal{O}({\rm TeV})$, the ensuing $\zeta$ lies in the ballpark
$\zeta\simeq 1$--$2$ (the precise value follows from the $g_{*S}$ evolution in the
two sectors).

\subsection{Temperature ratio \texorpdfstring{$\zeta\equiv T'/T$}{zeta = T'/T}: general expression and benchmarks}
\label{app:xi-table}

In our minimal secluded setup the late-time dark bath is dominated by the real
scalar \(H_p\) (mass \(m_{H_p}\)) and, depending on the benchmark, possibly by a
light dark gauge boson \(Z'\). The Dirac DM \(\chi\) (mass \(m_\chi\)) is
Boltzmann suppressed at \(T'\sim m_\chi/20\) and can be neglected in
\(g_{*S}^{\rm dark}(T')\) unless explicitly stated.

In order to estimate the value of $\zeta$ at the epoch relevant for DM
freeze--out, \(T\simeq m_\chi/20\), we use the following inputs:
\begin{itemize}
\item Visible sector: \(g_{*S}^{\rm vis}(T_{\rm dec})=106.75\),
      \(g_{*S}^{\rm vis}(T\simeq m_\chi/20)=86.25\).
\item Dark sector at decoupling: \(g_{*S}^{\rm dark}(T'_{\rm dec})=4.5\)
      (Dirac $\chi$ + real scalar $H_p$). If a dark gauge boson \(Z'\) is
      relativistic at decoupling, use \(g_{*S}^{\rm dark}(T'_{\rm dec})=7.5\).
\item Chemical potentials: we assume efficient number--changing in the dark bath
      (\(\mu_i=0\)) so that Eq.~\eqref{eq:xi_gstar_limit_app} applies directly.
      After dark chemical freeze-out (\(\mu_i>0\)) one must use the general form
      \eqref{eq:xi-entropy-master}, which typically yields a \emph{smaller}
      \(\zeta\) at fixed parameters.
\end{itemize}

Table~\ref{tab:xi-grid} summarizes representative values of $\zeta$ as a
function of \(m_{H_p}\) and \(m_\chi\) in the minimal case (no light \(Z'\) at
late times), assuming \(\mu_i=0\).
The qualitative trends are:
(i) for fixed \(m_\chi\), the heavier \(m_{H_p}\), the larger \(\zeta\);
(ii) for fixed \(m_{H_p}\), the larger \(m_\chi\), the smaller \(\zeta\).
Across the broad range relevant here, \(\zeta\) typically lies in the
\(1.1\)--\(1.9\) band.

\begin{table}[t]
\centering
\renewcommand{\arraystretch}{1.12}
\caption{Representative values of \(\zeta\equiv T'/T\) at
\(T\simeq m_\chi/20\), obtained by solving
Eq.~\eqref{eq:xi_gstar_limit_app} self-consistently with \(T'=\zeta T\),
\(g_{*S}^{\rm vis}(T)/g_{*S}^{\rm vis}(T_{\rm dec})=86.25/106.75\),
\(g_{*S}^{\rm dark}(T'_{\rm dec})=4.5\) (Dirac \(\chi\) + real scalar \(H_p\)),
and exact scalar entropy for \(H_p\) (no light \(Z'\) at late times,
\(\mu_i=0\)).}
\label{tab:xi-grid}
\begin{tabular}{c|ccccc}
\hline
\multirow{2}{*}{$m_{H_p}$ [GeV]} & \multicolumn{5}{c}{$\zeta$ for $m_\chi$ [GeV]} \\
 & 10 & 30 & 100 & 300 & 1000 \\
\hline
 1   & 1.296 & 1.238 & 1.236 & 1.236 & 1.236 \\
 3   & 1.601 & 1.296 & 1.238 & 1.236 & 1.236 \\
 10  & 1.872 & 1.639 & 1.296 & 1.238 & 1.236 \\
 30  & 1.894 & 1.872 & 1.601 & 1.296 & 1.238 \\
 100 & 1.895 & 1.894 & 1.872 & 1.639 & 1.296 \\
\hline
\end{tabular}
\end{table}

\subsection{Dark freeze--out with $T'=\zeta T$}
\label{sec:2T-framework}

After visible--dark decoupling, the portal remains too feeble to re--equilibrate
the sectors. The dark bath stays internally thermalized at temperature
$T'=\zeta T$ through fast secluded reactions. For the Dirac dark matter particle
$\chi$ (mass $m_\chi$) annihilating into a lighter real scalar $H_p$
($m_{H_p}<m_\chi$) via the Yukawa interaction $y_p\bar\chi\chi H_p$, the
dominant number--changing channel is
\begin{equation}
\chi\bar\chi \to H_p H_p.
\end{equation}

We work with the visible--entropy yield
$Y_\chi\equiv n_\chi/s_{\rm vis}$, using
$s_{\rm vis}(T)=\frac{2\pi^2}{45}g_{*S}^{\rm vis}(T)\,T^3$, and define
\begin{equation}
x\equiv \frac{m_\chi}{T},\qquad
x'\equiv \frac{m_\chi}{T'}=\frac{x}{\zeta}.
\end{equation}
During radiation domination, the Hubble rate receives contributions from the
radiation in both sectors,
\begin{equation}
H(T)
=\sqrt{\frac{8\pi G}{3}\,\big(\rho_{\rm vis}(T)+\rho_{\rm dark}(T')\big)}
=\frac{\pi}{\sqrt{90}}\frac{T^{2}}{M_{\rm Pl}}\,
\Big[g_{*}^{\rm vis}(T)+g_{*}^{\rm dark}(T')\,\zeta^{4}\Big]^{1/2}
\equiv \frac{1.66\,\sqrt{g_*^{H}(T,T')}}{M_{\rm Pl}}\,T^2,
\label{eq:H_general_app}
\end{equation}
where
\begin{equation}
g_*^{H}(T,T') \equiv g_*^{\rm vis}(T)
+\zeta^{4}\,g^{\rm dark}_{\ast}(T')
\end{equation}
is the energy degrees of freedom combination entering $H$.

The Boltzmann equation for $Y_\chi$ reads
\begin{equation}
\frac{dY_\chi}{dx}
= - \frac{s_{\rm vis}(T)}{x\,H(T)}\,
\langle\sigma v\rangle_{\chi\bar\chi\to H_pH_p}(x')\,
\Big[Y_\chi^2-(Y_\chi^{\rm eq}(x'))^2\Big],
\label{eq:BE_2T_app}
\end{equation}
with $Y_\chi^{\rm eq}(x')=n_\chi^{\rm eq}(T')/s_{\rm vis}(T)$ and where the
annihilation rate must be evaluated at the \emph{dark} temperature $T'$.

In App.~\ref{app:xschichiHpHp} we report the full calculation of the thermal
average for the process $\chi \bar{\chi}\to H_p H_p$ that dominates the relic
density:
\begin{equation}
\langle\sigma v\rangle_{\chi\bar\chi\to H_pH_p}(x')
=\frac{6\,b(r)}{x'}
=\frac{y_p^4}{64\pi m_\chi^2}\;\mathcal{S}_p(r)\;\frac{6}{x'}\,.
\label{eq:sigmav_thermal_app}
\end{equation}
For a Majorana $\chi$ the result is multiplied by $1/2$ due to identical initial
states. The kinematic limits are $\mathcal{S}_p(0)=1$ (light mediator) and
$\mathcal{S}_p(r)\propto (1-r)^{1/2}$ near threshold ($r\to 1$).

The freeze--out condition equates the \emph{dark} annihilation rate to the
\emph{visible} Hubble rate,
\begin{equation}
n_\chi^{\rm eq}(T')\,\langle\sigma v\rangle(T') \;\simeq\; H(T),
\label{eq:RDcond_app}
\end{equation}
with $n_\chi^{\rm eq}(T')\simeq g_\chi\big(\tfrac{m_\chi T'}{2\pi}\big)^{3/2}e^{-x'}$
in the non--relativistic regime. Using Eq.~\eqref{eq:sigmav_thermal_app} one
obtains the standard logarithmic solution
\begin{equation}
x_f'
\simeq
\ln\!\Big(\mathcal{A}\,y_p^4\Big)
-\frac{1}{2}\,\ln\!\Big[\ln\!\big(\mathcal{A}\,y_p^4\big)\Big],
\label{eq:xf_iter_app}
\end{equation}
with
\begin{equation}
\mathcal{A}\equiv
\frac{c_f\,g_\chi\,m_\chi M_{\rm Pl}}{1.66\,(2\pi)^{3/2}}\,
\frac{\zeta^2}{\sqrt{g_*^{H}(T_f,T_f')}}\,
\frac{6\,\mathcal{S}_p(r)}{64\pi m_\chi^2}\,,
\qquad
T_f\equiv \frac{m_\chi}{x_f},\quad T_f'\equiv \frac{m_\chi}{x_f'}.
\label{eq:A_prefactor_app}
\end{equation}
Here $c_f=\mathcal{O}(1)$ encodes the usual matching convention. For
$m_\chi\sim\mathrm{GeV}$ and $y_p\sim 0.1$--1 one finds
$x_f'\in[20,30]$, as in the standard $p$--wave freeze--out.

\begin{figure}
\includegraphics[width=0.49\linewidth]{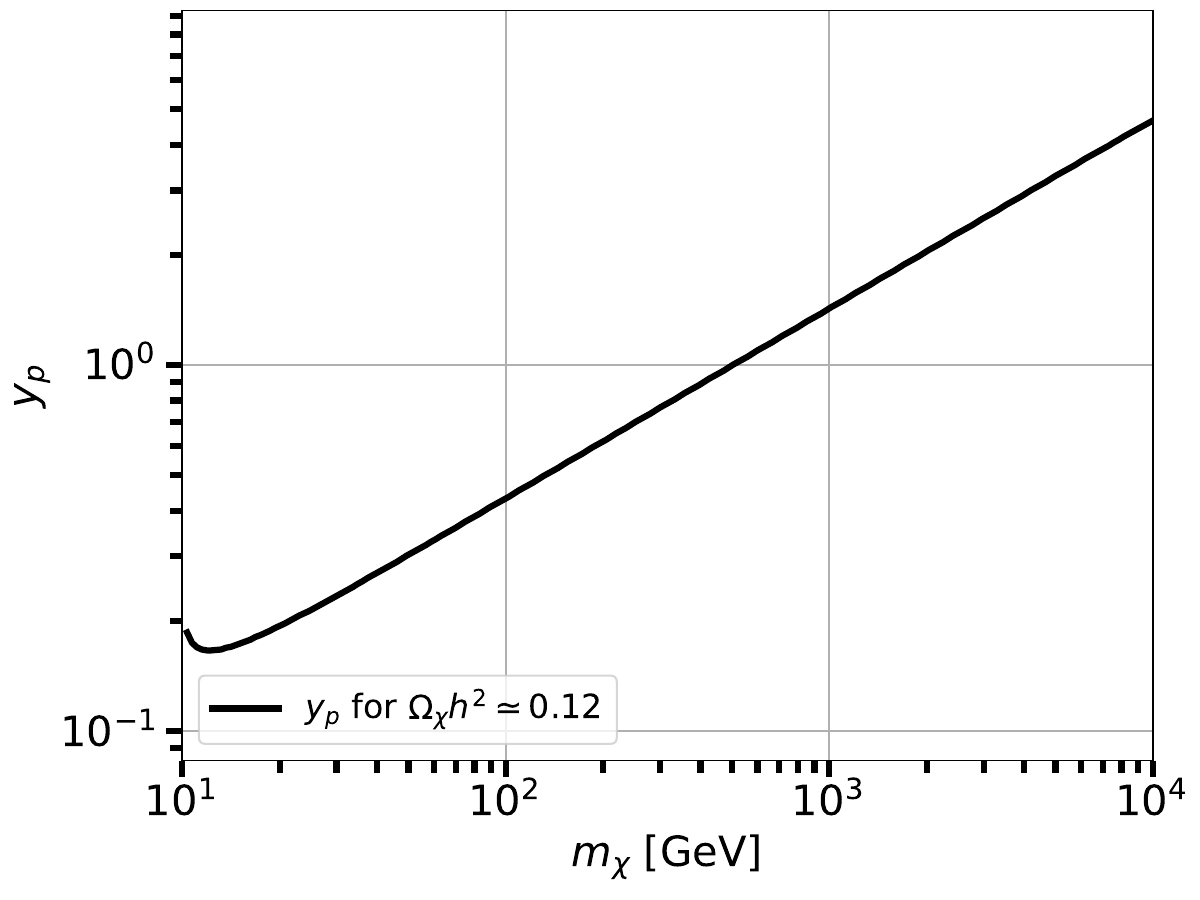}
\caption{Values of $y_p$ for which DM achieves the observed relic density.}
\label{fig:yprelic}
\end{figure}

The solution of Eq.~\eqref{eq:BE_2T_app} can be written in terms of the
annihilation integral in the two--temperature case,
\begin{equation}
J_{2T}\equiv \int_{x_f}^{\infty}\frac{\langle\sigma v\rangle(x')}{x^2}\,dx
=\frac{1}{\zeta}\left[\frac{a}{x_f'}+\frac{3\,b}{x_f'^2}\right].
\label{eq:J2T_app}
\end{equation}
Using for the annihilation cross section Eq.~\eqref{eq:sigmav_thermal_app} (pure
$p$--wave, $a=0$) we find
\begin{equation}
J_{2T}
=\frac{1}{\zeta}\;\frac{3\,y_p^4}{64\pi m_\chi^2}\;
\frac{\mathcal{S}_p(r)}{x_f'^2}\,.
\label{eq:J2T_p_app}
\end{equation}
Defining
\begin{equation}
\lambda_{\rm eff}\equiv
0.264\;\frac{g_{*S}^{\rm vis}(T_f)}{\sqrt{g_*^{H}(T_f,T_f')}}\;M_{\rm Pl}\,m_\chi,
\end{equation}
the asymptotic yield and relic density are
\begin{equation}
Y_\infty=\frac{1}{\lambda_{\rm eff}\,J_{2T}},
\qquad
\Omega_\chi h^2
\simeq \Big(1.05\times10^9\;{\rm GeV}^{-1}\Big)\,
\frac{\sqrt{g_*^{H}(T_f,T_f')}}{g_{*S}^{\rm vis}(T_f)}\;
\frac{1}{M_{\rm Pl}}\;\frac{1}{J_{2T}}.
\label{eq:Omega_master_app}
\end{equation}
Specializing to $p$--wave only and inserting Eq.~\eqref{eq:J2T_p_app},
\begin{equation}
\Omega_\chi h^2
\simeq
\Big(1.05\times10^9\;{\rm GeV}^{-1}\Big)\,
\frac{\sqrt{g_*^{H}(T_f,T_f')}}{g_{*S}^{\rm vis}(T_f)}\;
\frac{\zeta \, x_f'^2}{M_{\rm Pl}}\;
\frac{64\pi m_\chi^2}{3\,y_p^4\,\mathcal{S}_p(r)}\,.
\label{eq:Omega_pwave_app}
\end{equation}
Inverting Eq.~\eqref{eq:Omega_pwave_app} for the coupling,
\begin{equation}
y_p
=\left[
\Big(1.05\times10^9\;{\rm GeV}^{-1}\Big)\,
\frac{\sqrt{g_*^{H}(T_f,T_f')}}{g_{*S}^{\rm vis}(T_f)}\;
\frac{\zeta \,x_f'^2}{M_{\rm Pl}}\;
\frac{64\pi m_\chi^2}{3\,\Omega_\chi h^2\,\mathcal{S}_p(r)}\;
\right]^{1/4}.
\label{eq:yp_invert_app}
\end{equation}
A useful numerical estimate is
\begin{equation}
y_p \simeq 0.43
\left(\frac{x_f'}{25}\right)^{1/2}
\left(\frac{m_\chi}{100~\mathrm{GeV}}\right)^{1/2}
\left(\frac{g_{*S}^{\rm vis}}{86.25}\right)^{-1/4}
\left(\frac{g_*^{H}}{100}\right)^{1/8}.
\label{eq:yprelic}
\end{equation}
Therefore, a value of $y_p\sim \mathcal{O}(0.1$--$1)$ is needed to reach the
observed DM relic density. 
We show in Fig.~\ref{fig:yprelic} the values of $y_p$ for which DM achieves the observed relic density. Equivalently, a handy heuristic for the
\emph{required} thermal rate is
\begin{equation}
\langle\sigma v\rangle_{\rm req}
\;\approx\;
\frac{2\times 10^{-26}\ \mathrm{cm^{3}\,s^{-1}}}{\zeta}\,
\left[1+\mathcal{O}\!\left(\frac{g_*^{\rm dark}}{g_*^{\rm vis}}\,\zeta^{4}\right)\right],
\label{eq:sigmav_req_app}
\end{equation}
i.e.\ a hotter (colder) dark bath with $\zeta>1$ ($\zeta<1$) requires a
\emph{smaller} (\emph{larger}) annihilation cross section to reproduce the
observed $\Omega_{\rm DM} h^2$.

\section{Spin--independent nuclear cross section}
\label{app:SI-cross}

Spin--independent scattering of the DM particle $\chi$ off a nucleon
$N=p,n$ proceeds via $t$--channel exchange of the CP--even scalars $H$
and $H_p$. In our model the relevant couplings are
\begin{equation}
g_{\chi\chi H}   = \frac{y_p}{\sqrt{2}}\sin\alpha,\qquad
g_{\chi\chi H_p} = \frac{y_p}{\sqrt{2}}\cos\alpha,\qquad
g_{HNN}          = \frac{m_N}{v_h}\,f_N\cos\alpha,\qquad
g_{H_p NN}       = -\,\frac{m_N}{v_h}\,f_N\sin\alpha,
\end{equation}
where $f_N\simeq 0.30$ is the nucleon scalar form factor, defined through the
matrix element
\begin{equation}
m_N\,f_N \;=\; \sum_{q=u,d,s}\langle N |\, m_q\,\bar q q \,| N\rangle.
\end{equation}
The spin--independent cross section on a nucleon is
\begin{equation}
\sigma^{\rm SI}_{\chi N}
=\frac{\mu_N^2}{\pi}\left[
\sum_{i=H,H_p}
\frac{g_{\chi\chi h_i}\,g_{h_i NN}}{m_{h_i}^2}
\right]^2
= \frac{\mu_N^2 m_N^2 f_N^2 y_p^2}{2\pi v_h^2}\,
\sin^2\!\alpha\,\cos^2\!\alpha\,
\left(\frac{1}{m_H^2}-\frac{1}{m_{H_p}^2}\right)^2,
\label{eq:sigmaSI-def-app}
\end{equation}
with $\mu_N$ the $\chi$--$N$ reduced mass. This expression exhibits a 
``blind spot'' when $m_H\simeq m_{H_p}$, where the two scalar contributions
nearly cancel. In what follows we focus instead on the \emph{small-mixing}
regime relevant for secluded DM, $\sin\alpha\ll 1$.

For small mixing angles, the $H$--$H_p$ mixing is controlled by the mixed
quartic $\kappa$ in the potential
$V(\Phi,R)\supset \kappa\,(\Phi^\dagger\Phi)(R^\dagger R)$, where $\kappa$ is
connected to the mixing angle through
\begin{equation}
\tan 2\alpha = \frac{2\kappa v_h v_r}{m_{H_p}^2 - m_H^2}\,.
\label{eq:tan2alpha-portal-app}
\end{equation}
In the small-mixing limit ($|\alpha|\ll 1$),
\begin{equation}
\sin\alpha\,\cos\alpha
= \frac{1}{2}\sin 2\alpha
\simeq \frac{1}{2}\tan 2\alpha
\simeq \frac{\kappa v_h v_r}{m_{H_p}^2 - m_H^2}\,.
\label{eq:sincos-kappa-app}
\end{equation}
Inserting Eq.~\eqref{eq:sincos-kappa-app} into Eq.~\eqref{eq:sigmaSI-def-app},
one finds the compact form
\begin{equation}
\sigma^{\rm SI}_{\chi N}
=
\frac{\mu_N^2 m_N^2 f_N^2 y_p^2}{2\pi}\;
\frac{\kappa^2 v_r^2}{m_H^4 m_{H_p}^4}.
\label{eq:sigmaSI-kappa-compact}
\end{equation}

The portal coupling $\kappa$ is generated radiatively by the neutrino sector if
it vanishes (or is highly suppressed) at tree level. In the
Casas--Ibarra aligned limit ($O=\mathbb{1}$) and for three generations the
loop-induced portal coupling is (see Eq.~\ref{eq:kloop-simple})
\begin{equation}
\kappa_{\rm loop}(\mu)
=
-\sum_{I=1}^3
\frac{y_{N,I}^2\,M_I}{8\pi^2 v_h^2}\;m_{\nu,I}.
\label{eq:kloop-simplee}
\end{equation}
Inserting Eq.~\eqref{eq:kloop-simplee} into
Eq.~\eqref{eq:sigmaSI-kappa-compact} and \emph{eliminating} $v_r$ in favour of
$M_N$ and $y_N$, one finds
\begin{equation}
\sigma^{\rm SI}_{\chi N}
=
\frac{\mu_N^2 m_N^2 f_N^2 y_p^2}{64\pi^5 m_H^4 m_{H_p}^4 v_h^4}
\left(
\sum_{I=1}^3 y_{N,I}^2\,m_{\nu,I}^2\,M_I^4
\right).
\label{eq:sigmaSI-3gen}
\end{equation}
This form has \emph{no} residual dependence on $v_r$: the cross section is
fully expressed in terms of $y_N$, $M_N$, the light neutrino masses $m_{\nu,I}$,
and known SM parameters. A numerical form of
Eq.~\eqref{eq:sigmaSI-3gen} is
\begin{equation}
\sigma^{\rm SI}_{\chi N}
\simeq
3.9\times 10^{-57}~\mathrm{cm}^2\;
\left(\frac{m_\chi}{m_\chi + m_N}\right)^2
\left(\frac{y_p}{1}\right)^2
\left(\frac{10~\mathrm{GeV}}{m_{H_p}}\right)^4
\sum_{I=1}^3
\left(\frac{y_{N,I}}{1}\right)^2
\left(\frac{m_{\nu,I}}{0.05~\mathrm{eV}}\right)^2
\left(\frac{M_{N,I}}{10\,\mathrm{TeV}}\right)^4.
\label{eq:sigmaSI-numeric-3gen}
\end{equation}

\begin{figure}
\includegraphics[width=0.49\linewidth]{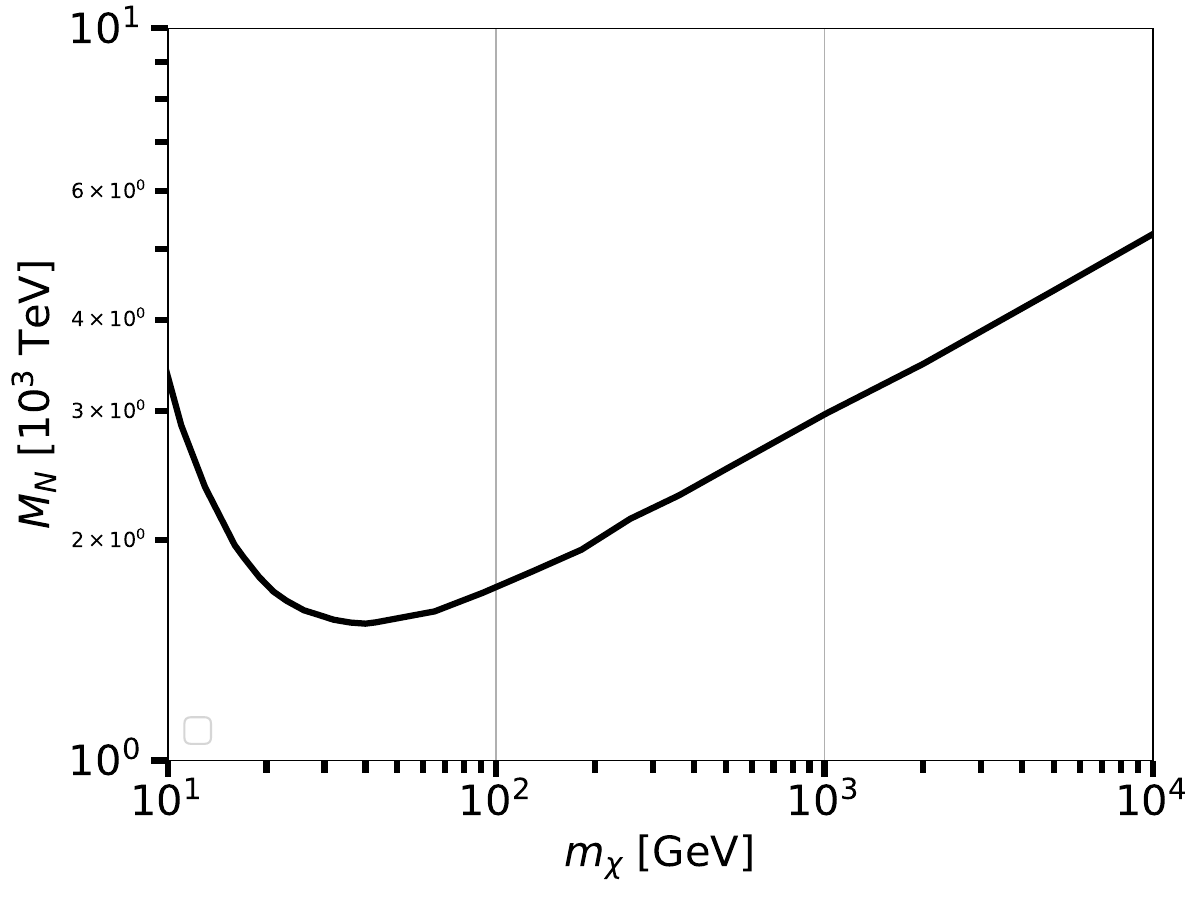}
\caption{
Upper limit on the heavy–neutrino mass $M_N$ as a function of the DM mass $m_\chi$, obtained by applying the spin-independent limits from Ref.~\cite{LZ:2024zvo} to our model (see Eq.~\ref{eq:sigmaSI-numeric-3gen}).
}
\label{fig:MNlimits}
\end{figure}

Just to provide a rough estimate of the bounds on the model from the latest LZ
spin--independent limit, we can take $m_\chi\simeq 100~\mathrm{GeV}$, for which
$\sigma_{\rm LZ} \simeq 3.5 \times 10^{-48}~\mathrm{cm}^2$. Using the
single--generation version of Eq.~\eqref{eq:sigmaSI-numeric-3gen} (with
$m_{H_p}=10~\mathrm{GeV}$) and solving for $M_N$, one finds an upper limit for
the heavy-neutrino mass of the order of
\begin{equation}
M_N
\lesssim
1.7\times 10^3~\mathrm{TeV}\;
\left(\frac{1}{y_p y_N}\right)^{1/2}
\left(\frac{0.05~\mathrm{eV}}{m_\nu}\right)^{1/2}
\left(\frac{m_{H_p}}{10~\mathrm{GeV}}\right).
\label{eq:MNmax-LZ}
\end{equation}
For $y_p\sim 1$ (the value required to reach the correct relic abundance, see
Eq.~\ref{eq:yprelic}), $y_N\sim 1$, $m_\nu\sim 0.05~\mathrm{eV}$ and
$m_{H_p}\sim 10~\mathrm{GeV}$, the loop-induced scalar portal is safely
consistent with LZ bounds for heavy-neutrino masses up to the few-$10^3~\mathrm{TeV}$ scale. 
In Fig.~\ref{fig:MNlimits} we show the upper limits for $M_N$ obtained when changing $m_\chi$. We can see that the upper limits remain in the few PeV range even changing the DM mass.
In other words, within this radiative portal
setup, current direct-detection limits do not impose a strong upper bound on
$M_N$; the requirement $|\tan\alpha|\lesssim 10^{-3}$ is easily satisfied
throughout the parameter space of interest.

\bibliographystyle{apsrev4-2}
\bibliography{main}

\end{document}